\documentclass[a4paper,fleqn,usenatbib]{mnras}

\usepackage{newtxtext,newtxmath}

\usepackage[T1]{fontenc}
\usepackage{ae,aecompl}

\usepackage{graphicx}	
\usepackage{amsmath}	
\usepackage{xcolor}           
\usepackage{xfrac}
\usepackage{gensymb}
\usepackage{pdflscape}
\usepackage{longtable}

\usepackage{subfig}

\title[WISDOM: The $\Delta V_\mathrm{CO}{-}M_\mathrm{BH}$ relation]{WISDOM project - VI. Exploring the relation between supermassive black hole mass and galaxy rotation with molecular gas}

\author[Mark D. Smith et al.]{
Mark D. Smith,$^{1}$\thanks{E-mail: mark.smith@physics.ox.ac.uk}
Martin Bureau,$^{1,2}$
Timothy A. Davis,$^{3}$
Michele Cappellari,$^{1}$
\newauthor{
Lijie Liu,$^{1}$
Kyoko Onishi,$^{4,5,6}$
Satoru Iguchi,$^{5,6}$
Eve V. North,$^{3}$
and Marc Sarzi$^{7}$}
\\
$^{1}$Sub-department of Astrophysics, Department of Physics, University of Oxford, Denys Wilkinson Building, Keble Road, Oxford, OX1 3RH, UK\\
$^{2}$Yonsei Frontier Lab and Department of Astronomy, Yonsei University, 50 Yonsei-ro, Seodaemun-gu, Seoul 03722, Republic of Korea\\
$^{3}$School of Physics \& Astronomy, Cardiff University, Queens Buildings, The Parade, Cardiff, CF24 3AA, UK\\
$^{4}$Research Center for Space and Cosmic Evolution, Ehime University, Matsuyama, Ehime, 790-8577, Japan\\
$^{5}$Department of Astronomical Science, SOKENDAI (The Graduate University of Advanced Studies), Mitaka, Tokyo 181-8588, Japan\\
$^{6}$National Astronomical Observatory of Japan, National Institutes of Natural Sciences, Mitaka, Tokyo, 181-8588, Japan\\
$^{7}$Armagh Observatory and Planetarium, College Hill, Armagh, BT61 DG, UK
}

\date{Accepted 2020 October 15. Received 2020 October 5; in original form 2020 February 11}

\pubyear{2020}

\begin{document}
\label{firstpage}
\pagerange{\pageref{firstpage}--\pageref{lastpage}}
\maketitle

\begin{abstract}
Empirical correlations between the masses of supermassive black holes (SMBHs) and properties of their host galaxies are well-established. Among these is the correlation with the flat rotation velocity of each galaxy measured either at a large radius in its rotation curve or via a spatially-integrated emission line width. We propose here the use of the de-projected integrated CO emission line width as an alternative tracer of this rotation velocity, that has already been shown useful for the Tully-Fisher (luminosity-rotation velocity) relation. We investigate the correlation between CO line widths and SMBH masses for two samples of galaxies with dynamical SMBH mass measurements, with respectively spatially-resolved and unresolved CO observations. The tightest correlation is found using the resolved sample of 24 galaxies as $\log (M_\mathrm{BH}/\mathrm{M_\odot})=(7.5\pm0.1)+(8.5\pm0.9)[\log(W_\mathrm{50}/\sin i \,\mathrm{km\,s}^{-1})-2.7]$, where $M_\mathrm{BH}$ is the central SMBH mass, $W_{50}$ the full-width at half-maximum of a double-horned emission line profile, and $i$ the inclination of the CO disc. This relation has a total scatter of $0.6\,$dex, comparable to those of other SMBH mass correlations, and dominated by the intrinsic scatter of $0.5\,$dex. A tight correlation is also found between the de-projected CO line widths and the stellar velocity dispersions averaged within one effective radius. We apply our correlation to the COLD GASS sample to estimate the local SMBH mass function.
\end{abstract}




\section{Introduction}
\label{sec_introduction}
Supermassive black holes (SMBHs) are found in the centres of almost all massive galaxies. They are now believed to play an instrumental role in the evolution of their hosts, a conclusion drawn from the tight correlations, spanning multiple orders of magnitude, between the SMBH masses and large-scale host properties (for reviews, see e.g. \citealt{Kormendy+2013ARAA51.511, Graham2016ASSL418.263}). This is surprising given that, in almost all galaxies following these correlations, the SMBH only dominates the gravitational potential on very small spatial scales (${\lesssim}100\,$pc), and it has a negligible gravitational influence on the scales on which the host properties are measured. Nevertheless, these tight correlations have been used to argue that each SMBH coevolves with its host, invoking mechanisms such as active galactic nucleus (AGN) feedback \citep[e.g.][]{Croton+2006MNRAS365.11,Bower+2006MNRAS370.645, Vogelsberger+2014MNRAS444.1518}, major merger-enhanced accretion \citep[e.g.][]{Sanders+1988ApJ325.74, Cattaneo+1999MNRAS308.77, DiMatteo+2005Nature433.604}, and simple merger averaging \citep[e.g.][]{Peng2007ApJ671.1098, Hirschmann+2010MNRAS407.1016, Jahnke+2011ApJ734.92}. The relative importance of these mechanisms remains under dispute \citep[e.g.][]{Kormendy+2013ARAA51.511,Simmons+2017MNRAS470.1559}.

The millimetre-Wave Interferometric Survey of Dark Object Masses (WISDOM) project aims to use the high angular resolutions achievable with the latest large millimetre/sub-millimetre interferometers to probe the environments around SMBHs. Principally, the survey is measuring SMBH masses using molecular gas kinematics, as the interferometers can resolve spatial scales dominated by the SMBHs \citep{Davis+2013Nature494.328, Onishi+2017MNRAS468.4663, Davis+2018MNRAS473.3818, Smith+2019MNRAS485.4359, Smith+2020MNRASsubmitted2, North+2019MNRAS490.319}. Other groups have also used the same method to constrain SMBH masses \citep[e.g.][]{Onishi+2015ApJ806.39, Barth+2016ApJL822.28, Barth+2016ApJ823.51, Boizelle+2019ApJ881.10, Ruffa+2019MNRAS489.3739, Nagai+2019ApJ883.193, Nguyen+2020ApJ892.68}. In addition to measuring SMBH masses, the same data can be used to investigate the physical state of the molecular gas and study giant molecular clouds \citep{Utomo+2015ApJ803.16, Liu+2020MNRASinrev}. In this paper, we use relatively low-resolution CO observations to investigate a correlation between deprojected integrated CO line widths and SMBH masses, that we propose could be a useful estimator of SMBH masses.

\subsection{Dynamical signatures of SMBH--baryonic matter correlations}
\label{ssec_DynamicalSMBHBaryon}
The tightest known correlations are those between the SMBH mass and properties of the host galaxy's bulge - stellar velocity dispersion \citep[$\sigma_\ast$; e.g.][]{Ferrarese+2000ApJL539.9, Gebhardt+2000ApJL539.13}, mass or luminosity \citep[$M_\mathrm{bulge}$ and $L_\mathrm{bulge}$; e.g.][]{Kormendy+1995ARAA33.581, Magorrian+1998AJ115.2285, Marconi+2003ApJL589.21, Haring+2004ApJL604.89}. These host quantities are measures of the stellar mass-dominated central potential of the bulge.

The $M_\mathrm{BH}{-}\sigma_\ast$ relation has traditionally been viewed as the tightest correlation, with an intrinsic scatter of only ${\approx}0.3\,$dex \citep[e.g.][]{Gebhardt+2000ApJL539.13, Beifiori+2012MNRAS419.2497, Saglia+2016ApJ818.47}. The velocity dispersions used for these investigations are usually taken from spectra integrated within 1 effective radius ($R_\mathrm{e}$), although the effects of finite instrumental apertures can affect this scale.

Correlations also exist between the SMBH mass and other properties of the host galaxy. One of the simplest such properties is the total stellar mass ($M_\mathrm{\ast,tot}$; \citealt{Davis+2018ApJ869.113}). Although originally the total stellar mass (and the disc component in late-type galaxies, LTGs) was thought not to correlate with SMBH mass \citep{Kormendy+2001AIPC586.363, Kormendy+2011Nature469.374}, recent works have indicated there is a correlation, albeit one weaker than $M_\mathrm{BH}{-}\sigma_\ast$ \citep[e.g.][]{Beifiori+2012MNRAS419.2497, Lasker+2014ApJ780.70, Savorgnan+2016ApJS222.10, Mutlu-Pakdil+2018MNRAS474.2594} with $0.66\,$dex scatter \citep{Davis+2018ApJ869.113}. This allows total stellar mass (or luminosity) to be invoked as a convenient proxy for SMBH mass where the dynamical $\sigma_\ast$ is unavailable or hard to measure.

The total stellar mass/luminosity can also be linked to a dynamical quantity via the shape of the rotation curve, as indicated for spirals by the Tully-Fisher relation \citep{TullyFisher1977AandA54.661}. Such rotation curves are observed from the line-of-sight projected velocities of (rotating) dynamical tracers (e.g. \citealt{Pease1918PNAS4.21, Burbidge+1959ApJ130.739, Rubin+1970ApJ159.379}, see \citealt{Sofue+2001ARAandA39.137} for a review), and have been extensively used to study the structure of galaxies. Each galactic rotation curve is a probe of the gravitational potential, from the SMBH-dominated central region \citep[occasionally spatially-resolvable with modern high-resolution observations; e.g.][]{Greenhill+1995AandA304.21, Gao+2017ApJ834.52, North+2019MNRAS490.319}, through a stellar-mass dominated regime, to outer parts dominated by a putative dark halo. A rotation curve can be decomposed into contributions from these components of the galactic potential \citep[e.g.][]{Martinsson+2013AandA557.131}. Disc galaxies often exhibit rotation curves that are (almost) flat over most of the stellar-dominated regimes and into the halo-dominated regimes, providing evidence for the `disc-halo conspiracy' \citep[e.g.][]{vanAlbada+1986RSPTA320.447, Williams+2009MNRAS400.1665}. The rotation curves of some elliptical and lenticular galaxies peak within $1\,R_\mathrm{e}$ before declining with increasing radius, some then flattening in the outer parts \cite[e.g.][]{Noordermeer+2007MNRAS376.1513, Cappellari+2013MNRAS432.1709}.

\subsection{CO kinematics as a baryonic matter tracer}
\label{ssec_COlinesTraceBaryons}
Both $M_\mathrm{\ast,tot}$ and $\sigma_\ast$ are widely known to correlate with the spatially-integrated line width of neutral hydrogen (\ion{H}{I}; e.g. \citealt{Whitmore+1979ApJ234.68, Courteau+2014RvMP86.47, Serra+2016MNRAS460.1382}) . However, each has also been linked to the line width of CO. Throughout this discussion we refer to the spatially-integrated width of an emission line as $\Delta V$, with a subscript denoting the emitting atom/molecule.

\subsubsection{CO Tully-Fisher relation}
\label{ssec_COTFR}
The Tully-Fisher relation (TFR), relating the asymptotic velocity of a rotationally-supported disc to its host galaxy's absolute magnitude, is widely used to measure distances to extragalactic sources. For the last two decades, since initial proposals by \cite{DickeyKazes1992ApJ393.530} and \cite{Sofue1992PASJ44.231}, millimetre-wave emission from CO molecules has been used as an alternative to radio \ion{H}{I} or optical H$\alpha$ emission to trace the asymptotic rotation velocity \citep[e.g.][]{Ho2007ApJ669.821, Davis+2011MNRAS414.968, Davis+2016MNRAS455.214, Tiley+2016MNRAS461.3494, Topal+2018MNRAS479.3319,Tiley+2019MNRAS482.2166}, including in early-type galaxies (ETGs) where such rotationally-supported CO discs are still reasonably common \citep[e.g.][]{Davis+2013MNRAS429.534}. 

Although CO discs do not usually extend as far as neutral hydrogen discs, they have been found to extend to the flat part of the rotation curve in ${\approx}70\%$ of CO-rich ETGs \citep{Davis+2013MNRAS429.534}, and they appear to in most LTGs (e.g. \citealt{Leung+2018MNRAS477.254, Levy+2018ApJ860.92}). The potential at the radii probed by CO, however, is still dominated by the stellar component \citep{Cappellari+2013MNRAS432.1709}, and thus CO gas does not trace the same halo-dominated potential as \ion{H}{I}. We discuss this in further detail in Section \ref{ssec_COrotationCurves}.

Nevertheless, \cite{Davis+2019ApJ877.64} combined the CO TFR of \cite{Tiley+2019MNRAS482.2166} with the $M_\mathrm{BH}{-}M_\mathrm{\ast, tot}$ correlation of \cite{Davis+2018ApJ869.113} to yield a prediction of $M_\mathrm{BH} \propto \Delta V_\mathrm{CO}^{12.2\pm2.1}$. For spiral galaxies, the disc-halo conspiracy enables the replacement of the CO line widths in the \cite{Tiley+2019MNRAS482.2166} TFR with those of \ion{H}{I}. The same argument cannot be made for ETGs, as we discuss in Section \ref{ssec_haloes}. In a sample of 48 spiral galaxies with dynamically-measured SMBH masses and \ion{H}{I} line widths, \cite{Davis+2019ApJ877.64} obtained a relation of $M_\mathrm{BH}\propto \Delta V_\mathrm{\ion{H}{I}}^{10.62\pm1.37}$, consistent with the prediction. Notably, both the predicted and the observed relation are substantially steeper than found in earlier works \cite[e.g.][]{Beifiori+2012MNRAS419.2497, Sabra+2015ApJ803.5}. 

\subsubsection{\ion{H}{I} and CO line widths correlate with $\sigma_\ast$}
\label{ssec_lineWidthBaryons}
A correlation between spatially-integrated \ion{H}{I} emission line widths and $\sigma_\ast$ was initially suggested for disc galaxies by \cite{Whitmore+1979ApJ234.68} and \cite{Whitmore+1981ApJ250.43}, where a constant $\Delta V_\ion{H}{I}/\sigma_\ast {\approx}1.7$ was observed in local S0 and spiral galaxies. This ratio has since been shown to vary substantially with redshift and morphology \citep[e.g.][]{Cresci+2009ApJ697.115}. The most accurate $\Delta V_\ion{H}{I}{-}\sigma_\ast$ correlation for ETGs was obtained using spatially-resolved rotation curves from the ATLAS$^\mathrm{3D}$ survey. \cite{Serra+2016MNRAS460.1382} found a linear relation, $\Delta V_\ion{H}{I}/\sigma_\ast=1.33$, and 12$\%$ total scatter consistent with the measurement errors. We must once more emphasise that these correlations physically correspond to a correlation between a halo property and a property measured on baryon-dominated scales, discussed further in Section \ref{ssec_haloes}.

In addition to that of large-scale \ion{H}{I}, other emission line widths on smaller spatial scales have been proposed to potentially correlate with $\sigma_\ast$. \cite{NelsonWhittle1999ASPC182.53} showed for a sample of Seyfert galaxies that the full-width at half-maximum (FWHM) of the nuclear [\ion{O}{III}] emission correlates with its $\sigma_\ast$ when the latter is measured on a similar scale, with $\Delta V_{[\ion{O}{III}]}/\sigma_\ast=2.35$. \cite{Nelson2000ApJL544.91} later adopted this relation to use the [\ion{O}{III}] FWHM as a proxy for $\sigma_\ast$ to investigate the $M_\mathrm{BH}{-}\sigma_\ast$ relation among active galaxies, for which there were few $\sigma_\ast$ measurements available. 

\cite{Shields+2006ApJ641.683} proposed the use of the CO line width as a proxy for $\sigma_\ast$ to study whether the $M_\mathrm{BH}{-}\sigma_\ast$ relation was already in place in a sample of quasars at redshifts $2 < z < 6$ (since [\ion{O}{III}] is out of the optical window at these redshifts), assuming similarly that $\Delta V_\mathrm{CO}/\sigma_\ast=2.35$. Although \cite{Shields+2006ApJ641.683} thereby concluded that $M_\mathrm{BH}{-}\sigma_\ast$ does not hold at high redshifts, a subsequent analysis by \cite{Wu2007ApJ657.177} showed that by de-projecting $\Delta V_\mathrm{CO}$ and assuming the quasars were viewed almost face-on, the CO line widths could be brought into agreement with the local $M_\mathrm{BH}{-}\sigma_\ast$ relation. In his analysis of Seyfert galaxies, \cite{Wu2007ApJ657.177} found a steeper $\Delta V_\mathrm{CO}{-}\sigma_\ast$ correlation than \cite{Shields+2006ApJ641.683} had assumed, $\Delta V_\mathrm{CO}/(\sin i\,\mathrm{km\,s^{-1}}) = (-67.16\pm 80.18) \,+\,(3.62\pm0.68)(\sigma_\ast / \mathrm{km\,s^{-1}})$, where $i$ is the inclination of the CO disc. Although this relation still has substantial scatter, it dramatically outperforms the simpler approximation of \cite{Shields+2006ApJ641.683}. 

\subsection{What part of the rotation curve do spatially-integrated CO line widths trace?}
\label{ssec_COrotationCurves}
The spectral profile of a broad spatially-integrated emission line is dominated by Doppler-broadening, and thus tells us about the kinematics of the galaxy rather than the physical state of its emitting gas. Thus both \ion{H}{I} and CO line widths trace the rotation curve. However, as hinted at above, these emission lines originate from very different regions of the galaxy. \ion{H}{I} emission primarily traces the outer parts, where the rotation curve is typically very flat and the potential dominated by non-luminous matter. 

In contrast, CO emission typically extends to only ${\approx}0.5\,R_\mathrm{e}$ \citep{Davis+2013MNRAS429.534}, emphatically in baryon-dominated regions. It thus traces the depth of the central, stellar, potential in a manner similar to $\sigma_\ast$. As previously discussed (Section \ref{ssec_COTFR}), in LTGs this distinction is made smaller by the disc-halo conspiracy. However, this conspiracy is not known to hold in ETGs \citep{Young+2008ApJ676.317, Cappellari+2013MNRAS432.1709}. Furthermore, ETG rotation curves are not ubiquitously flat in their outer parts, with recent kinematic modelling indicating that the rotation curves often decline (from some maximum) with increasing radius, such that the asymptotic velocities traced by \ion{H}{I} are typically ${\approx}25\%$ lower than those found by central tracers at $0.2\,R_\mathrm{e}$ \citep{Serra+2016MNRAS460.1382}, as previously found in early-type disc galaxies \citep{Noordermeer+2007MNRAS376.1513}. This indicates the existence of an inner maximum in the rotation curve (hereafter $V_\mathrm{max}$). The ATLAS$^\mathrm{3D}$ survey constructed Jeans Anisotropic Models (JAMs) of ETGs, and from these generated a rotation curve for each object. \cite{Cappellari+2013MNRAS432.1709} showed that both the outer asymptotic circular velocity and $V_\mathrm{max}$ correlate tightly with $\sigma_\ast$, the latter with an intrinsic scatter of only $7\%$.

However, if one simply measures the width of an emission line, it is not trivial to identify which of these two velocities (asymptotic circular velocity or $V_\mathrm{max}$) is traced by CO emission \citep[e.g.][]{Noordermeer+2007MNRAS381.1463}. \cite{Lavezzi+1997AJ114.2437} have argued that the shapes of the emission line profiles can be used to select those that reach beyond $V_\mathrm{max}$, i.e. the profiles must be `boxy' or `sharp-edged'. With different specific implementations, this criterion is now widely used in the absence of spatially-resolved emission \citep[e.g.][]{Davis+2011MNRAS414.968, Tiley+2016MNRAS461.3494, Topal+2018MNRAS479.3319}.

However, even when $V_\mathrm{max}$ has not been reached such profiles can also be produced by sharply-truncated discs. Modern interferometric observations have revealed populations of (potentially low-surface brightness) CO discs in ETGs that are truncated at the edge of the associated circumnuclear dust discs (e.g. \citealt{Barth+2016ApJ823.51, Boizelle+2017ApJ845.170, Davis+2018MNRAS473.3818, Boizelle+2019ApJ881.10}). It is worth noting that some of these galaxies have nearly-flat rotation curves within the $V_\mathrm{max}$ defined by luminous mass models, due to the contribution of the central SMBHs, also leading to double-horned profiles, but spatially-resolved observations are nevertheless essential to determine which scales the CO emission probes.

\subsection{Rotation curves, haloes and SMBHs}
\label{ssec_haloes}
Following the discovery of the first SMBH mass-host property correlations, studies began to investigate whether the underlying coevolution was with the bulge, or whether a more fundamental correlation existed with another structural component. Particular interest revolved around the halo mass, with the asymptotic value of the rotation curve invoked as a suitable observable proxy. Initial work by \cite{Ferrarese2002ApJ578.90} and later \cite{Pizzella+2005ApJ631.785} appeared to show a non-linear relationship between the SMBH and (dark) halo masses. However, this was really a correlation between the rotation velocity and $\sigma_\ast$ of each galaxy, and it relied on invoking the $M_\mathrm{BH}{-}\sigma_\ast$ relation and the assumption that the asymptotic rotation velocity measured correlates with the (dark) halo mass.

The later analysis of \cite{Kormendy+2011Nature469.377} showed that there was no correlation unless the galaxy also hosted a classical bulge - thus the apparent correlation was merely an `indirect result of the rotation curve conspiracy'. Moreover, they argued that a loose correlation between these parameters cannot be taken to imply coevolution - on the simple principle that a larger galaxy contains larger structural components. \cite{Kormendy+2011Nature469.374} went on to challenge the assumption that $\sigma_\ast$ closely traces $M_\mathrm{BH}$ in pseudobulges, and hence the inference that haloes drive the growth of SMBHs.

Nevertheless, interest in these correlations has not subsided, not least because although weak correlations should not be taken as implying co-evolution, they can enable simple observable proxies to be used as estimators of SMBH masses. \cite{Davis+2018ApJ869.113}, for instance, suggested that their $M_\mathrm{\ast,tot}$ correlation is `beneficial for estimating $M_\mathrm{BH}$ from pipeline data or at higher redshift, conditions that are not ideal for the isolation of the bulge'.

However, we must be cautious. In Section \ref{ssec_lineWidthBaryons} we argued that the spatially-integrated width of a CO line traces the baryonic component of a galaxy, since CO emission does not extend to halo-dominated scales. Moreover, since it is unclear whether the disc-halo conspiracy holds in lenticulars (or an analogous relation in ellipticals), we argue that the discussion in this paper of a correlation between CO line widths and SMBH masses should not be taken as implying anything about SMBH-halo coevolution.

\subsection{A CO perspective on rotation curve correlations}
The discussion in Sections \ref{ssec_DynamicalSMBHBaryon} and \ref{ssec_COlinesTraceBaryons} implies that it is reasonable to investigate a correlation between the deprojected, spatially-integrated CO line widths of galaxies, tracing their rotation curves within the baryon-dominated regions, and their SMBH masses. Such a correlation may prove an alternative to the $M_\mathrm{BH}{-}\sigma_\ast$ relation. To this end, we fit double-horned emission line profiles to new and archival CO observations of galaxies to measure their line widths. We then show that these CO line widths correlate sufficiently well with the SMBH masses to be used as proxies to estimate SMBH masses. We also contribute to the extensive literature on CO line width correlations with $\sigma_\ast$, by finding a reasonably tight correlation.

This paper exploits three recent improvements: (1) we use only the most robust SMBH masses measured dynamically, (2) we derive our tightest relation with spatially-resolved, high spectral resolution and high signal-to-noise ratio (SNR) CO observations from the Atacama Large Millimetre/sub-millimetre Array (ALMA) and other interferometers (and show the negative impact of instead using unresolved single-dish spectra) and (3) we fit our spatially-integrated CO emission lines with `Gaussian double-peak' line profiles, that have been shown to recover well the intrinsic line widths \citep{Tiley+2016MNRAS461.3494}.

In Section \ref{sec_data}, we describe the observational data used in this study. This is accompanied by Appendix \ref{app_newObservations}, detailing new single-dish observations. Section \ref{sec_results} goes on to measure the CO line widths and explore potential correlations. In Section \ref{sec_discussion}, we discuss the conclusions that can be drawn from this study and the limitations of the sample used. We conclude in Section \ref{sec_conclusion}. 

\section{Data}
\label{sec_data}
SMBH masses have been measured in around 200 local galaxies over the last three decades, using a variety of dynamical tracers of the galaxies' central potentials. These results, as compiled in \cite{vdBosch2016ApJ831.134}, are used as the starting point for this work, to which a few more recent measurements are added. In addition to robust measurements, there are also a large number of upper limits, principally from ionised gas. In this work, we exclude such upper limits, leaving a parent sample of 196 galaxies with well-constrained SMBH masses. As most SMBH mass measurements require spatially-resolved tracers, these objects are typically well-studied local galaxies. Almost all have now been observed in CO using single-dish telescopes (though only 75 were detected), and 73 have been observed with interferometers (of which 58 were detected).

Investigating a correlation with CO line widths requires that we are able to recover these widths precisely. This is of particular importance since, while among massive galaxies the asymptotic rotation velocities span less than one order of magnitude ($100-500\,$km$\,$s$^{-1}$), the corresponding SMBH masses vary over four ($10^6-10^{10}\,\mathrm{M_\odot}$). Thus, a comparatively small uncertainty of a few tens of km$\,$s$^{-1}$ in a line width will translate to a large uncertainty in the predicted SMBH mass. The significance of this potentially large line width uncertainty is somewhat mitigated by the fact that even a dynamically-measured SMBH mass can exhibit a relatively large uncertainty.

The data required to constrain the line widths need to be of high quality to obtain robust line width measurements. Available observations comprise (intrinsically spatially-integrated) spectra from single-dish telescopes and spatially-resolved data cubes from either interferometers or mosaics of multiple pointings by single-dish telescopes. Although spatially-resolving the CO is formally unnecessary to measure spatially-integrated line widths, resolved observations offer multiple advantages. First, we can verify that the molecular gas discs are in ordered rotation, and thus that the line widths truly measure the depths of the potentials. Second, the achieved sensitivities are generally significantly higher, due to the arrays' larger total collecting areas compared to single dishes, and the similar (or longer) integration times. Third, the use of smaller individual antennae leads to array primary beams that are much larger than those of single-dish telescopes, avoiding the data potentially missing some emission at large galactic radii. Pointing errors for single-dish telescopes can also cause extended emission to be missed, leading to erroneously asymmetric (and potentially artificially narrowed) line profiles. Such pointing errors are trivially diagnosed with spatially-resolved images, and are in any case generally unimportant due to the large primary beams. This is particularly important for our sample, as the galaxies used are all sufficiently local (and thus extended on the sky) that their SMBH masses could be measured by resolving spatial scales on which the SMBHs dominate the potentials. These advantages are countered by the significantly higher complexity in obtaining, calibrating and imaging interferometric observations, although modern observing, data reduction and data analysis pipelines have now somewhat mitigated this challenge. For these reasons, observations that spatially resolve the CO discs are preferable.

Although spatially-resolved observations are to be preferred, such observations are not available for most galaxies. We therefore divide our objects into two samples, galaxies with respectively spatially-resolved and unresolved CO observations. We fit both samples using identical procedures, and in Section \ref{ssec_resUnresCompare} discuss the negative effects of using unresolved data.

From the parent sample of 196 candidate galaxies, we obtain CO spectra as described in Sections \ref{ssec_resolvedSample} and \ref{ssec_unresolvedSample}. The final samples are selected from these observations, applying the criteria discussed in Sections \ref{ssec_resolvedSample}, \ref{ssec_unresolvedSample} and \ref{ssec_linewidths}. Finally, a few galaxies that are clear outliers are excluded, and these are justified in Section \ref{ssec_outliers}. Table \ref{tab_sampleSelection} lists the number of galaxies that remain after each selection criterion is applied.

The CO spectra of the galaxies in the final samples, and all those taken in our new observations, are available online via Zenodo\footnote{\href{https://dx.doi.org/10.5281/zenodo.4067034}{https://dx.doi.org/10.5281/zenodo.4067034}}, excluding those that are already public from the Spectrographic Areal Unit for Research on Optical Nebulae (SAURON) and ATLAS$^\mathrm{3D}$ surveys \citep{Combes+2007MNRAS377.1795, Young+2011MNRAS414.940}\footnote{SAURON and ATLAS$^\mathrm{3D}$ spectra are hosted at \href{http://www-astro.physics.ox.ac.uk/atlas3d/}{http://www-astro.physics.ox.ac.uk/atlas3d/}}.

\begin{table}
	\centering
	\caption{Sample size after applying each selection criterion.}
	\label{tab_sampleSelection}
	\begin{tabular}{lccl} 
		\hline
		Selection criterion & Spatially- & Spatially- & Section\\
		&resolved & unresolved&\\
		& sample & sample&\\
		(1) & (2) & (3) & (4)\\
		\hline
		Observed & 73 & 162 &\ref{ssec_resolvedSample}/\ref{ssec_unresolvedSample} \\
		CO detected & 58 & 75 & \ref{ssec_resolvedSample}/\ref{ssec_unresolvedSample}\\
		Regular rotation$^\ast$ & 37 & -- & \ref{ssec_resolvedSample}\\
		Boxy profile & 29 & 53 & \ref{ssec_COrotationCurves}, \ref{ssec_resolvedSample}/\ref{ssec_unresolvedSample}\\
		Accepted fit & 27 & 24 & \ref{ssec_linewidths}\\
		Not omitted & 25 & 21 & \ref{ssec_outliers}\\
		\hline
	\end{tabular}
	\parbox{0.47\textwidth}{\textbf{Notes:} Column 1 lists each selection criterion applied to the samples, described in the main text. Column 2 lists the number of galaxies with spatially-resolved observations that remain after each criterion is applied, while Column 3 lists the corresponding number of galaxies with unresolved observations. Column 4 lists the section(s) in which each criterion is discussed. The criterion marked with a $^\ast$ only applies to the spatially-resolved sample.}
\end{table}

\subsection{Spatially-resolved CO observations}
\label{ssec_resolvedSample}
Data cubes (right ascension, declination, and velocity) are produced from either interferometric observations or by mosaicking multiple pointings of a single-dish telescope. We obtain such cubes from the ALMA archive, the Berkeley-Illinois-Maryland Association Survey of Nearby Galaxies\footnote{\href{https://ned.ipac.caltech.edu/level5/March02/SONG/SONG.html}{https://ned.ipac.caltech.edu/level5/March02/SONG/SONG.html}} \citep[BIMA-SONG;][]{Helfer+2003ApJS145.259} and the ATLAS$^\mathrm{3D}$ survey\footnote{\href{http://www-astro.physics.ox.ac.uk/atlas3d/}{http://www-astro.physics.ox.ac.uk/atlas3d/}} \citep{Alatalo+2013MNRAS432.1796}. ALMA observations have been calibrated and imaged automatically either by the ALMA pipeline or manually by ALMA Regional Centre staff, and the cubes used are those provided on the archive, with the exception of observations taken by our own WISDOM programme and its precursor studies, for which the data reduction and calibration (and the properties of the data cubes) are described in the associated papers \citep{Davis+2013Nature494.328,Onishi+2017MNRAS468.4663, Davis+2017MNRAS468.4675, Davis+2018MNRAS473.3818, Smith+2019MNRAS485.4359, North+2019MNRAS490.319, Smith+2020MNRASsubmitted2}. BIMA-SONG and ATLAS$^\mathrm{3D}$ data cubes are used as provided on the associated websites. We visualise each cube and manually select only those that appear to show overall rotation, leaving 37 galaxies.

We then convert each data cube into a spectrum by integrating each channel over the two spatial dimensions. The emission in any given channel typically extends over only a few pixels, with the remainder populated by noise. To simply sum all pixels together blindly would needlessly include all this noise in our sum, degrading the sensitivity of the resultant spectrum. We can do better by summing only the pixels contained within a mask encompassing all the emission.

Such a mask can be generated using the `smooth-masking' technique \citep{Dame2011arXiv1101.1499}, originally developed to make high-quality moment maps. Each cube is first smoothed spatially by the beam and Hanning-smoothed spectrally. Pixels with values exceeding a noise threshold in the smoothed cube are included in the mask; we use a threshold of $5$ times the rms noise measured in the original cube. The mask is then applied to the original, unsmoothed cube, and should encompass all high-surface brightness pixels, assumed to correspond to real emission, as well as small regions around them that may include lower surface brightness emission. An integrated spectrum is then produced by summing all pixels included in the mask.

The resulting spectra are of generally significantly higher quality than those obtained by single-dish telescopes. However, the uncertainty on the total flux in each channel must be considered carefully. In a normal single-dish spectrum, we can simply measure the rms noise in line-free channels, and assume that it is constant across the full bandpass. However, in integrated spectra derived from data cubes, the uncertainty in each channel is instead a function of the number of pixels included in the mask in that channel. We estimate this in each channel by assuming the noise to be Gaussian with standard deviation $\sigma_\mathrm{px}$. The sum of $N_\mathrm{mask}$ normal random variables is then $\sigma_\mathrm{px} \sqrt{N_\mathrm{mask}}$, where $N_\mathrm{mask}$ is the number of pixels in the mask in that channel. This formalism is only valid in channels where the mask is non-zero. 

In channels where the mask is zero (i.e. $N_\mathrm{mask}=0$), both the integrated flux and the estimated uncertainty would be zero. There may however be real emission below the detection threshold. We assume that such emission would be distributed on a spatial scale comparable to those in other channels, and adopt the mean number of pixels in the non-zero channels of the mask ($\langle N_\mathrm{mask}\rangle$) as the representative spatial scale over which undetected emission would be distributed. The uncertainty is thus given by $\sigma_\mathrm{void} = \sigma_\mathrm{px} \sqrt{\langle N_\mathrm{mask}\rangle}$, and it is therefore constant for line-free channels. 

The resulting spectra are all visually inspected, and those that do not have a boxy line profile are rejected. This can occur even when the dynamics appear to exhibit overall rotation, (e.g. if the CO distribution does not sample the velocity field well). Applying this criterion leaves 29 galaxies with spatially-resolved CO observations.

\subsection{Unresolved sample}
\label{ssec_unresolvedSample}
Line widths can also be measured directly from the spectra obtained in single-dish observations. Such observations are simpler than spatially-resolved observations, but typically have lower SNRs.  The other challenge with these spectra in local galaxies is the relatively small primary beams of the large telescopes used, that may not extend far enough into the galaxies to reach the flat parts of their rotation curves in single pointings. For these reasons, there are two concerns with the line widths obtained from the objects in this unresolved galaxy sample: the uncertainties are larger fractions of the channel width than those estimated for the resolved sample galaxies, and we cannot be certain whether the line widths measured encompass the full widths of the rotation curves. 

The literature contains single-dish CO observations in two forms. Most commonly, spectra are shown in figures only, with the quantitative information needed for other astronomers to use the data rarely available. We then use the public tool \texttt{GraphClick}\footnote{\href{http://www.arizona-software.ch/graphclick/}{http://www.arizona-software.ch/graphclick/}} to manually digitise these figures, obtaining flux measurements by interpolating from the axis scales.

These data are given variously in the antenna ($T_\mathrm{A}^\ast$), radiation ($T_\mathrm{R}^\ast$) and main beam ($T_\mathrm{mb}$) temperature scales. For the sake of homogeneity, we transform them all to the same flux density scale (Jy) as the spectra obtained for our resolved galaxy sample, using observatory-specific appropriate beam efficiencies listed in Table \ref{tab_telEff}. We assume that the emission is point-like and do not account for the spatially-varying responses of the telescopes. This would be valid only if the gas is centrally concentrated. As we have argued above, and discuss further in Section \ref{ssec_resUnresCompare}, the relatively small extents of the beams compared to the very large extents of these local galaxies imply this assumption may be invalid. With no \textit{a priori} information on the gas distributions, however, we cannot make more appropriate conversions. Moreover, since this is a spectrally-constant conversion, any change will not affect the line widths measured, the only quantities used in this paper. 

For all spectra except those of \cite{Maiolino+1997ApJ485.552} (for which a direct conversion from $T_\mathrm{R}^\ast$ is provided), we therefore first convert from $T_\mathrm{mb}$ or $T_\mathrm{R}^\ast$ to $T_\mathrm{A}^\ast$ using the efficiencies in Table \ref{tab_telEff}, and then convert from $T_\mathrm{A}^\ast$ to flux densities using
\begin{equation}
S\,\mathrm{(Jy)} = \frac{2k_\mathrm{B}}{\eta_\mathrm{A}}\,\frac{4}{\pi D^2}\,T_\mathrm{A}^\ast\,,
\end{equation}
where $k_\mathrm{B}$ is Boltzmann's constant, $\eta_\mathrm{A}$ the telescope aperture efficiency (relating the geometric area of the telescope dish to its effective area) and $D$ the telescope diameter.

In addition to observations published by other authors, we acquired new observations at the Institut de Radioastonomie Millim\'etrique (IRAM) 30-m telescope under programme 191-18 and at the Onsala Space Observatory (OSO) 20-m telescope under programme 2018-04a. Fifty-one galaxies were observed with the IRAM 30-m telescope, of which twenty-two were detected. Nine were observed at the OSO 20-m telescope, of which four were detected. These observations are described in detail in Appendices \ref{app_191-18} and \ref{app_2018-04a}, respectively. Associated noise estimates, integrated line fluxes and total molecular gas masses are listed in Tables \ref{tab_191-18Results} and \ref{tab_o2018-04aResults}, and the spectra and telescope beam extents are shown in Figures \ref{fig_191-18spectra} and \ref{fig_o2018-04a_spectra}, that are continued as Figures A3 and A4 in the supplemental material. 

Combining observations from the literature and from our programmes, there are unresolved CO detections of 75 of the galaxies in our parent sample. From these, those without a boxy line profile are rejected, leaving 53 galaxies. Estimates of the noise levels in all these spectra are obtained from the emission-free channels at either end of the CO lines, and are assumed to be spectrally constant.

\begin{table}
\begin{center}
\caption{Adopted conversions from literature units to flux densities, assuming point sources.}
\label{tab_telEff}
\begin{tabular}{lcl}
    \hline
    Telescope & CO line &  Conversions \\ 
    \hline
    IRAM 30-M & CO(1-0)   & $T_\mathrm{A}^\ast=0.78\,T_\mathrm{mb}$  \\
                     &             & $S_\mathrm{Jy} = 5.98\,T_\mathrm{A}^\ast$  \\
                     & CO(2-1)  & $T_\mathrm{A}^\ast=0.63\,T_\mathrm{mb}$  \\
                     &             & $S_\mathrm{Jy} = 7.73\,T_\mathrm{A}^\ast$ \\ 
    \hline
    NRAO 12-M & CO(1-0) & $T_\mathrm{R}^\ast=0.91\,T_\mathrm{mb}$  \\
                      &            & $S_\mathrm{Jy} = 35\,T_\mathrm{R}^\ast$  \\
    \hline
\end{tabular}
\end{center}
\end{table}

\section{Results}
\label{sec_results}
We discussed in Section \ref{ssec_COTFR} the role of the CO Tully-Fisher relation to interpret our proposed CO line width--SMBH mass correlation. However, different methods of measuring the CO line width from a spectrum have been proposed. The simplest scheme is simply to adopt the width at which the observed flux first falls below some fraction of the maximum \citep[e.g.][]{Davis+2011MNRAS414.968}. Using $20\%$ of the maximum flux appears to yield a tighter CO TFR correlation \citep{Ho2007ApJ669.821}, but $50\%$ would be preferable at low SNRs where a smaller fraction cannot be accurately determined. This approach can be particularly unreliable with very low SNR spectra, where although the line can be visually identified as a consistently-positive sequence of channels, the line edges are ill-defined. A profile fit to the line is therefore now generally preferred. This latter method also allows spectra with anomalous line shapes to be rejected when poorly fit by a suitably physically-motivated profile.

\subsection{Line-width measurements}
\label{ssec_linewidths}
\cite{Tiley+2016MNRAS461.3494} investigated appropriate choices of line profiles and determined that the `Gaussian double peak' profile, consisting of a quadratic function bounded by half-Gaussian wings, gave the most reliable line width measurements with least sensitivity to SNR and inclination. We therefore adopt the full-width at half-maximum (FWHM; equivalent to the $50\%$ criterion defined above) of such a profile as our measure of the line width, fitting each spectrum with the \cite{Tiley+2016MNRAS461.3494} function
\begin{equation}
f(v) = \begin{cases}
A_\mathrm{G} \times e^\frac{-[v-(v_0-w)]^2}{2\sigma^2} & v \leq v_0-w\\
A_\mathrm{C} + a(v-v_0)^2 & v_0-w \leq v \leq v_0+w\\
A_\mathrm{G} \times e^\frac{-[v-(v_0+w)]^2}{2\sigma^2} & v_0+w \leq v
\end{cases},
\label{eq_doubleGaussian}
\end{equation}
where $A_\mathrm{G}$ (the flux of each peak), $A_\mathrm{C}$ (the flux of the central extremum), $v_0$ (the velocity of the central extremum), $w$ (the velocity half-width of the quadratic function) and $\sigma$ (the velocity width of both half-Gaussian functions) are all free parameters, and $a$ is determined by the continuity conditions at $v_0\pm w$. The corresponding line width at the half-maximum is then given by 
\begin{equation}
W_\mathrm{50} = 2(w+\sigma\sqrt{2\ln2}) \;.
\label{eq_W50}
\end{equation} 
We note that in cases for which $A_\mathrm{C}>A_\mathrm{G}$, this equation will not yield the FWHM, yielding instead slightly broader line widths. We describe below that galaxies for which $A_\mathrm{C} > (3/2)A_\mathrm{G}$ (those that would be most affected by this effect) are in any case rejected for not being double-horned profiles, and in the few cases where this condition occurs in the remaining galaxies this broadening effect is negligible due to the sharp edges of the spectral lines and $A_\mathrm{C}$ being only very slightly greater than $A_\mathrm{G}$.

\begin{figure}
	\includegraphics[trim={0.5cm 1cm 1.5cm 1cm}, width=\columnwidth]{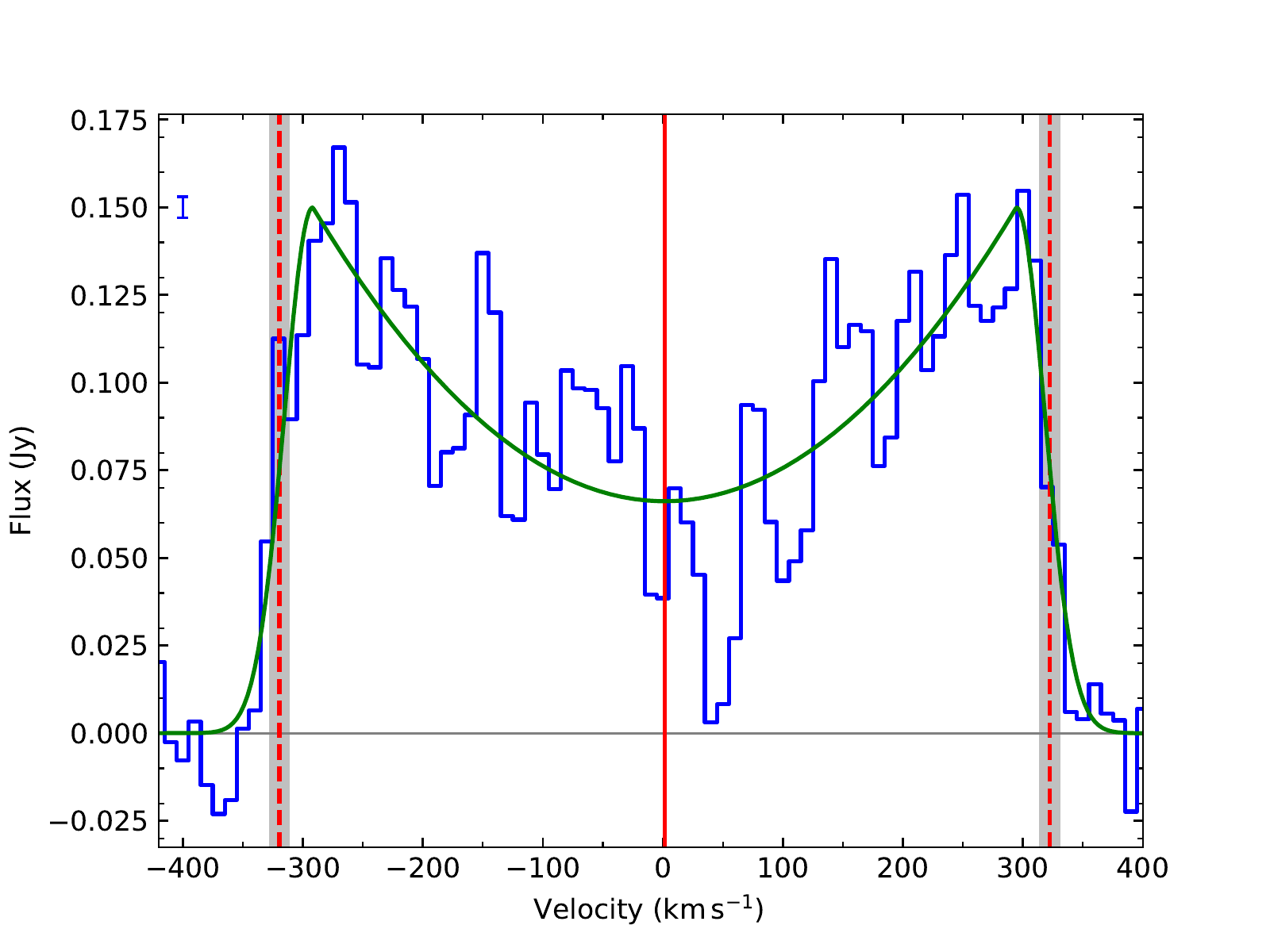}
	\caption{Example Gaussian double peak profile fit (green) overlaid on a spectrum synthesised from the CARMA data cube of NGC3665 (blue). The mean noise estimate ($\pm1\sigma$) is shown in the top-left corner. The red vertical lines are at $v_0$ (solid) and bound $W_{50}$ (dashed), defined in Equation \ref{eq_W50}, while the grey bands indicate the $67\%$ confidence interval in $W_\mathrm{50}$. The original data cube had already had the galaxy's systemic velocity subtracted. Fits to the other spectra of spatially-resolved galaxies are shown in Figure 9 in the supplemental material.}
    \label{fig_resolved_linefits}
\end{figure}

\begin{figure}
	\includegraphics[trim={0.5cm 1cm 1.5cm 1cm}, width=\columnwidth]{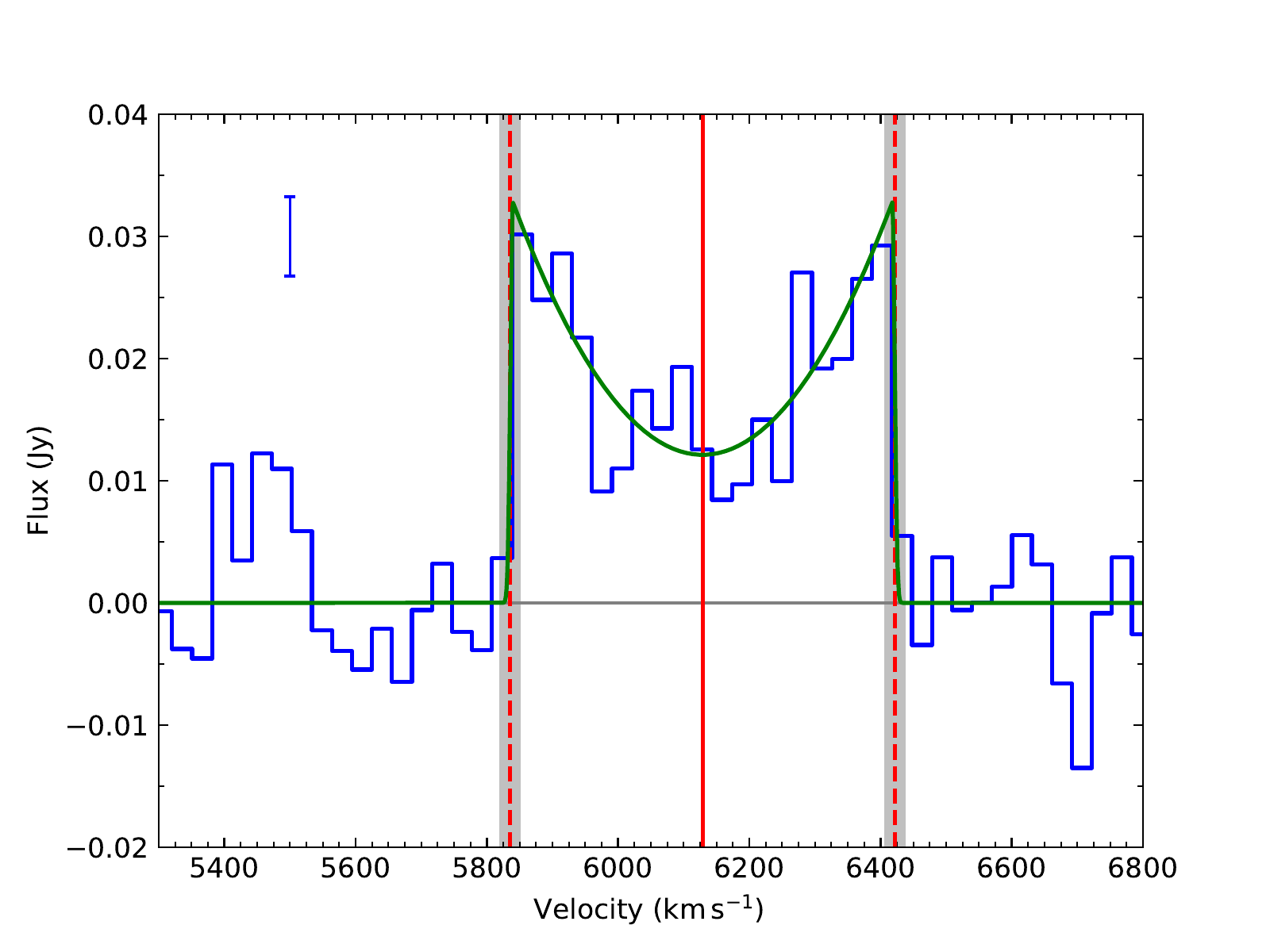}
    \caption{Example Gaussian double peak profile fit (green) overlaid on the IRAM 30-m spectrum of \mbox{NGC 1497}. The mean noise estimate ($\pm1\sigma$) is shown in the top-left corner. The red vertical lines are at $v_0$ (solid) and bound $W_{50}$ (dashed), defined in Equation \ref{eq_W50}, while the grey bands indicate the $67\%$ confidence interval in $W_\mathrm{50}$. The velocity axis is the observed velocity, and includes the galaxy's systemic velocity. Fits to the other spectra of unresolved galaxies are shown in Figure 10 in the supplemental material.}
    \label{fig_unresolved_linefits}
\end{figure}

The fits are performed using the Python package \texttt{lmfit}\footnote{\href{https://lmfit.github.io/lmfit-py/}{https://lmfit.github.io/lmfit-py/}}, minimising the chi-squared statistic:
\begin{equation}
\chi^2 = \sum_i\left(\frac{\mathrm{data}_i - \mathrm{model}_i}{\sigma_i}\right)^2 ,
\end{equation}
where $i$ denotes each velocity bin (i.e. channel) of the spectrum, and $\sigma_i$ is the uncertainty on the flux at each velocity bin as described in Section \ref{sec_data}. Each spectrum is fit 30 times, with initial conditions selected randomly from a uniform distribution within reasonable physical limits, to ensure a global minimum is found. The fit with the smallest reduced chi-square is then selected as the best-fitting solution. The associated line width and uncertainty is then estimated from the uncertainties on $\sigma$ and $w$ (determined by the \texttt{lmfit} routines) through Monte Carlo methods: ($\sigma$, $w$) pairs are generated as normal random variables with means given by the best-fitting ($\sigma$, $w$) and standard deviations obtained from the \texttt{lmfit}-derived uncertainties. The corresponding $W_\mathrm{50}$ are then calculated from Equation \ref{eq_W50}, and the median and standard deviation adopted as the best-fitting line width and uncertainty respectively.

We also investigated the systematics affecting our line width measurements. We generate 150 realisations of the best-fitting model with random normally-distributed noise of magnitude equal to the noise in the data added to each. Each realisation is then fit by the same Gaussian double peak profile using the parameters best-fitting the data as initial conditions. The standard deviations of the distributions of line widths from these fits are comparable to the uncertainties estimated by \texttt{lmfit} for the spectra in our unresolved sample, but they are much smaller (and also much smaller than the channel width) for our spatially-resolved sample due to the much higher signal-to-noise ratios. This suggests that reducing the noise in the spectra of the unresolved sample would yield smaller line width uncertainties, whereas adopting smaller channel widths will yield the greatest improvement for galaxies in the spatially-resolved sample.

Since not all profiles are well-reproduced by a Gaussian double peak profile, we also fit each profile with a simple Gaussian. We immediately reject a spectrum if either the Gaussian profile has a lower reduced $\chi^2$ or the Gaussian double peak profile has $A_\mathrm{C} > (3/2)A_\mathrm{G}$, the latter corresponding to the flux at the centre of the spectrum being at least $50\%$ higher than that of the bounding half-Gaussians, in which case one might more naturally fit the spectrum with a single Gaussian \citep[e.g.][]{Tiley+2016MNRAS461.3494}. Both criteria correspond to a violation of the `boxy' criterion outlined previously, and imply that the observed CO gas is unlikely to have reached the flat part of the rotation curve. 

Finally, every spectrum is manually inspected to ensure a good fit was achieved. This leaves final samples of 27 galaxies with spatially-resolved observations and 24 galaxies with unresolved observations that satisfy our selection criteria. These galaxies, and the associated line width measurements, inclinations and SMBH masses, are listed in Tables \ref{tab_resolvedData} and \ref{tab_unresolvedData}, respectively. Example profile fits to interferometric and single-dish observations are shown in Figures \ref{fig_resolved_linefits} and \ref{fig_unresolved_linefits} respectively, and all data and associated fits are shown in Figures 9 and 10 in the supplemental material.

\begin{table*}
	\centering
	\caption{CO data, best-fitting line widths and host galaxy properties for our spatially-resolved galaxies. All galaxies listed have a well-determined CO line width, but one (\mbox{NGC 5055}) is excluded from the final correlations. This omission is justified in Section \ref{ssec_outliers}.}
	\label{tab_resolvedData}
	\begin{tabular}{lccccccc} 
		\hline
		Name & T-type & CO transition & $W_{50}$ & Inclination & $\log(M_\mathrm{BH}/\mathrm{M}_\odot)$ & SMBH method & Notes\\
		           &          &             & (km$\,$s$^{-1}$) & ($\degree$) \\
		(1) & (2) & (3) & (4) & (5) & (6) & (7) & (8)\\
		\hline
		Circinus    & \phantom{-}3.3 & 1-0$^\mathrm{1\phantom{0}}$ & $337.0\pm\phantom{1}3.0$ & $78\pm\phantom{1}1^\mathrm{2\phantom{0}}\,$(M) & $6.06\pm0.10^\mathrm{3\phantom{0}}$ & masers & $M_\mathrm{BH}{-}\sigma_\ast$ outlier\\
		NGC 383   & -2.9 & 2-1$^\mathrm{4\phantom{0}}$ & $522.0\pm\phantom{1}3.7$ & $38\pm\phantom{1}4^\mathrm{4\phantom{0}}\,$(M) & $9.62\pm0.07^\mathrm{4\phantom{0}}$ & CO\\
		NGC 524  & -1.2 & 2-1$^\mathrm{5\phantom{0}}$ & $292.4\pm13.8$ & $20\pm\phantom{1}5^\mathrm{6\phantom{0}}\,$(D) & $8.60\pm0.30^\mathrm{5\phantom{0}}$ & CO\\
		NGC 1332 & -2.9 & 2-1$^\mathrm{7\phantom{0}}$ & $884.8\pm\phantom{1}4.8$ & $84\pm\phantom{1}1^\mathrm{8\phantom{0}}\,$(M) & $8.82\pm0.04^\mathrm{7\phantom{0}}$ & CO\\
		NGC 1386 & -0.7 & 1-0$^\mathrm{9\phantom{0}}$ & $373.6\pm\phantom{1}1.8$ & $68\pm\phantom{1}2^{10}\,$(D) & $6.07\pm0.29^{11}$ & masers & $M_\mathrm{BH}{-}\sigma_\ast$ outlier\\
		NGC 3081 & \phantom{-}0.0 & 2-1$^{12}$& $285.6\pm10.9$ & $46\pm10^{12}\,$(D) & $7.20\pm0.30^{13}$ & ionised gas\\
		NGC 3245 & -2.1 & 2-1$^\mathrm{14}$ & $579.2\pm\phantom{1}2.6$ & $63\pm\phantom{1}1^{15}\,$(K) & $8.28\pm0.11^{15}$ & ionised gas\\
		NGC 3258 & -4.3 & 2-1$^\mathrm{16}$ & $609.6\pm12.0$ & $45\pm\phantom{1}5^{16}\,$(M) & $9.35\pm0.01^{17}$ & CO\\
		NGC 3504 & \phantom{-}2.1 & 2-1$^\mathrm{18}$ & $215.4\pm10.4$ & $62\pm\phantom{1}3^{18}\,$(M) & $7.01\pm0.07^{18}$ & CO & Omitted\\
		NGC 3557 & -4.9 & 2-1$^\mathrm{19}$ & $431.3\pm19.1$ & $56\pm\phantom{1}1^{19}\,$(M) & $8.85\pm0.01^{19}$ & CO\\
		NGC 3607 & -3.2 & 2-1$^{20}$& $542.6\pm\phantom{1}8.9$ & $48\pm\phantom{1}6^{21}\,$(D) & $8.14\pm0.16^{22}$ & stars\\
		NGC 3627 & \phantom{-}3.1 & 1-0$^{23}$ & $389.1\pm13.1$ & $61\pm\phantom{1}1^{24}\,$(D) & $6.93\pm0.05^{25}$ & stars\\
		NGC 3665 & -2.1 & 2-1$^{26}$ & $641.9\pm15.4$ & $70\pm\phantom{ 1}1^{26}\,$(M) & $8.76\pm0.10^{26}$ & CO\\
		NGC 4258 & \phantom{-}4.0 & 1-0$^{23}$ & $388.1\pm20.4$ & $68\pm\phantom{1}4^{27}\,$(B) & $7.58\pm0.03^{28}$ & masers\\
		NGC 4303 & \phantom{-}4.0 & 2-1$^{29}$ & $121.0\pm\phantom{1}6.9$ & $21\pm10^{30}\,$(M) & $6.51\pm0.74^{31}$ & ionised gas\\ 
		NGC 4429 & -0.8 & 1-0$^{32}$ & $543.9\pm11.1$ & $\phantom{\!\!\!}66.8\pm0.2^{33}\,$(M) & $8.18\pm0.08^{34}$ & CO\\
		NGC 4459 & -1.6 & 1-0$^{34}$ & $382.8\pm14.5$ & $46\pm\phantom{1}2^{35}\,$(D) & $7.84\pm0.09^{36}$ & ionised gas\\
		NGC 4526 & -1.9 & 2-1$^{37}$ & $692.3\pm\phantom{1}5.9$ & $79\pm\phantom{1}3^{37}\,$(M) & $8.65\pm0.12^{37}$ & CO\\
		NGC 4697 & -4.5 & 2-1$^{38}$ & $435.7\pm\phantom{1}2.8$ & $76\pm\phantom{1}1^{38}\,$(M) & $8.11\pm0.06^{38}$ & CO\\  
		NGC 4736 & \phantom{-}2.3 & 1-0$^{23}$ & $236.5\pm\phantom{1}5.9$ & $35\pm10^{39}\,$(K) & $6.78\pm0.12^{40}$ & stars & FP outlier\\
		NGC 4826 & \phantom{-}2.2 & 1-0$^{23}$ & $327.8\pm\phantom{1}3.1$ & $60\pm\phantom{1}3^{21}\,$(D) & $6.05\pm0.12^{40}$ & stars & FP outlier\\
		NGC 5005 & \phantom{-}4.0 & 1-0$^{23}$ & $574.3\pm15.6$ & $69\pm\phantom{1}5^{10}\,$(D) & $8.27\pm0.23^{13}$ & ionised gas\\
		NGC 5055 & \phantom{-}4.0 & 1-0$^{23}$ & $363.2\pm\phantom{1}4.7$ & $59\pm\phantom{1}2^{10}\,$(D) & $8.70\pm0.90^{41}$ & ionised gas &  $M_\mathrm{BH}{-}\sigma_\ast$ outlier; omitted\\
		NGC 5248 & \phantom{-}4.0 & 1-0$^{23}$ & $241.6\pm\phantom{1}5.9$ & $54\pm\phantom{1}4^{21}\,$(D) & $6.30\pm0.38^{13}$ & ionised gas \\
		NGC 6861 & -2.7 & 2-1$^\mathrm{16}$ & $995.9\pm\phantom{1}6.2$ & $71\pm\phantom{1}5^{21}\,$(D) & $9.30\pm0.08^{42}$ & stars \\
		NGC 7052 & -4.9 & 2-1$^{43}$ & $712.7\pm10.2$ & $75\pm\phantom{1}5^{43}\,$(M) & $9.41\pm0.05^{43}$ & ionised gas\\ 
		NGC 7331 & \phantom{-}3.9 & 1-0$^{23}$ & $528.5\pm\phantom{1}8.2$ & $70\pm\phantom{1}4^{27}\,$(B) & $8.02\pm0.18^{13}$ & ionised gas\\
		\hline
	\end{tabular}
	\parbox{0.9\textwidth}{\textbf{Notes:} Column 1 lists the name of each galaxy contained in the final sample of spatially-resolved galaxies. The morphological classification on the numerical Hubble scale from \href{http://leda.univ-lyon1.fr}{HyperLEDA} is listed in Column 2. Spatially-resolved observations of the CO transition listed in Column 3 were integrated within a mask to obtain a spectrum. Column 4 lists the (line-of-sight projected) line width and associated uncertainty measured from a Gaussian double peak line profile fit to this spectrum. Column 5 lists the inclination of the CO disc and in parentheses the method used to measure it (D~-~dust morphology, M~-~molecular gas morphology/kinematics, B~-~\textit{B}-band apparent flattening, K~-~other kinematics). Column 6 lists the dynamically-measured SMBH mass using the tracer cited in Column 7. Column 8 contains other notes about certain galaxies. Footnotes in column 3 indicate the source of the CO observations, and those in columns 6 and 7 the source of the measurement, as follows. \textbf{References:} (1)~\cite{Zschaechner+2016ApJ832.142}, (2)~\cite{Curran+1998AandA338.863}, (3)~\cite{Greenhill+2003ApJ590.162}, (4)~\cite{North+2019MNRAS490.319}, (5)~\cite{Smith+2019MNRAS485.4359},  (6)~\cite{Cappellari+2006MNRAS366.1126}, (7)~\cite{Barth+2016ApJL822.28}, (8)~\cite{Barth+2016ApJ823.51}, (9)~\cite{Zabel+2019MNRAS483.2251}, (10) this work, from a \textit{HST} WFPC2 F606W image, (11)~\cite{Braatz+1997AAS19110402}, (12)~\cite{Ramakrishnan+2019MNRAS487.444}, (13)~\cite{Beifiori+2012MNRAS419.2497}, (14)~\href{https://almascience.eso.org/asax/?result_view=observation&projectCode=2017.1.00301.S&sourceNameResolver=NGC3245}{ADS/JAO.ALMA\#2017.1.00301.S}, (15)~\cite{Barth+2001ApJ546.205}, (16)~\cite{Boizelle+2017ApJ845.170}, (17)~\cite{Boizelle+2019ApJ881.10}, (18)~\cite{Nguyen+2020ApJ892.68}, (19)~\cite{Ruffa+2019MNRAS489.3739}, (20)~\href{https://almascience.eso.org/asax/?result_view=observation&projectCode=2015.1.00598.S&sourceNameResolver=NGC3607}{ADS/JAO.ALMA\#2015.1.00598.S}, (21)~this work, from a \textit{HST} WFPC2 F814W image, (22)~\cite{Gultekin+2009ApJ695.1577}, (23)~\cite{Helfer+2003ApJS145.259}, (24)~\cite{Casasola+2011AandA527.92},  (25)~\cite{Saglia+2016ApJ818.47}, (26)~\cite{Onishi+2017MNRAS468.4663}, (27)~\href{http://leda.univ-lyon1.fr}{HyperLEDA}, (28)~\cite{Herrnstein+2005ApJ629.719}, (29)~\cite{Sun+2018ApJ860.172}, (30)~\cite{Schinnerer+2002ApJ575.826}, (31)~\cite{Pastorini+2007AandA469.405}, (32)~\cite{Alatalo+2013MNRAS432.1796}, (33)~\cite{Davis+2018MNRAS473.3818}, (34)~\cite{Davis+2017MNRAS464.453},  (35)~\cite{Young+2008ApJ676.317}, (36)~\cite{Sarzi+2001ApJ550.65}, (37)~\cite{Davis+2013Nature494.328}, (38)~\cite{Davis+2017MNRAS468.4675}, (39)~\cite{Bosma+1977AandA57.373}, (40)~\cite{Kormendy+2011Nature469.374}, (41)~\cite{BlaisOuellette+2004AandA420.147}, (42)~\cite{Rusli+2013AJ146.45} and (43)~\cite{Smith+2020MNRASsubmitted2}.}
\end{table*}

\begin{table*}
	\centering
	\caption{CO data, best-fitting line widths and host galaxy properties for our spatially-unresolved galaxies. All galaxies listed have well-determined CO line widths, but a few are excluded from the final correlations. These omissions are justified in Section \ref{ssec_outliers}.}
	\label{tab_unresolvedData}
	\begin{tabular}{lccccccc} 
		\hline
		Name & T-type & CO transition & $W_{50}$ & Inclination & $\log(M_\mathrm{BH}/\mathrm{M}_\odot)$ & SMBH method & Notes\\
		           &                 &      & (km$\,$s$^{-1}$) & ($\degree$) \\
		(1) & (2) & (3) & (4) & (5) & (6) & (7) & (8)\\
		\hline
		3C120 & -1.7 & 1-0$^{1\phantom{1}}$ & $526.3\pm38.1$ & $65\pm\phantom{1}5^{2\phantom{1}}\,$(B) & $7.73\pm0.15^{3\phantom{1}}$ & reverberation\\
		Ark 120 & -5.0 & 1-0$^{4\phantom{1}}$ & $371.5\pm30.9$ & $49\pm\phantom{1}5^{5\phantom{1}}\,$(D) & $8.05\pm0.17^{6\phantom{1}}$ & reverberation \\ 
		Mrk 590 & \phantom{-}1.0 & 1-0$^{6\phantom{1}}$ & $223.9\pm\phantom{1}7.5$ & $26\pm\phantom{1}8^{2\phantom{1}}\,$(B) & $7.55\pm0.18^{8\phantom{1}}$ & reverberation \\ 
		NGC 383 & -2.9 & 1-0$^{9\phantom{1}}$ & $534.0\pm15.4$ & $38\pm\phantom{1}4^{10}\,$(M) & $9.62\pm0.07^{10}$ & CO \\
		NGC 524 & -1.2 & 1-0$^{11}$ & $310.2\pm54.5$ & $20\pm\phantom{1}5^{12}\,$(D) & $8.60\pm0.30^{13}$ & CO\\
		NGC 541 & -3.6 & 2-1$^{9\phantom{1}}$ & $271.0\pm43.2$ & $32\pm\phantom{1}5^{14}\,$(D) & $8.59\pm0.34^{15}$ & ionised gas & Omitted\\
		NGC 1068 & \phantom{-}3.0 & 1-0$^{16}$ & $288.8\pm24.2$ & $35\pm\phantom{1}8^{2\phantom{1}}\,$(B) & $6.92\pm0.25^{17}$ & masers\\
		NGC 1497 & -2.0 & 1-0$^{4\phantom{1}}$ & $587.0\pm28.2$ & $85\pm\phantom{1}5^{18}\,$(M) & $8.63\pm0.19^{15}$ & ionised gas\\
		NGC 1667 & \phantom{-}5.0 & 1-0$^{16}$ & $408.6\pm17.7$ & $40\pm\phantom{1}8^{2\phantom{1}}\,$(B) & $8.20\pm0.23^{15}$ & ionised gas\\
		NGC 1961 & \phantom{-}4.2 & 1-0$^{4\phantom{1}}$ & $429.5\pm\phantom{1}8.0$ & $46\pm\phantom{1}7^{2\phantom{1}}\,$(B) & $8.29\pm0.34^{15}$ & ionised gas\\
		NGC 2273 & \phantom{-}0.9 & 1-0$^{19}$ & $348.8\pm43.3$ & $58\pm\phantom{1}4^{2\phantom{1}}\,$(B) & $6.93\pm0.04^{20}$ & masers\\
		NGC 2911 & -2.0 & 1-0$^{4\phantom{1}}$ & $549.5\pm41.1$ & $63\pm\phantom{1}5^{2\phantom{1}}\,$(B) & $9.09\pm0.29^{15}$ & ionised gas\\
		NGC 3384 & -2.6 & 2-1$^{21}$ & $172.0\pm24.8$ & $62\pm\phantom{1}5^{22}\,$(K) & $7.03\pm0.21^{23}$ & stars & Omitted\\
		NGC 3665 & -2.1 & 1-0$^{11}$ & $632.7\pm41.5$ & $70\pm\phantom{1}1^{24}\,$(M) & $8.76\pm0.09^{24}$ & CO\\
		NGC 3862 & -4.8 & 1-0$^{9\phantom{1}}$ & $212.1\pm13.7$ & $15\pm15^{25\phantom{1}}\,$(D) & $8.41\pm0.37^{15}$ & ionised gas & Omitted\\ 
		NGC 4388 & \phantom{-}2.8 &1-0$^{16}$ & $304.6\pm26.9$ & $90^2\phantom{\pm0001\!}\,$(B) & $6.86\pm0.04^{20}$ & masers\\
		NGC 4429 & -0.8 & 1-0$^{11}$ & $521.7\pm14.8$ & $\phantom{1\!\!\!\!\!\!}66.8\pm0.2^{26}\,$(M) & $8.17\pm0.03^{26}$ & CO\\
		NGC 4459 & -1.6 & 1-0$^{27}$ & $387.3\pm50.2$ & $46\pm\phantom{1}2^{12}\,$(D) & $7.84\pm0.09^{28}$ & ionised gas\\
		NGC 4486 & -4.3 & 1-0$^{9\phantom{1}}$ & $421.3\pm47.1$ & $42\pm\phantom{1}5^{29}\,$(K) & $9.58\pm0.10^{30}$ & ionised gas\\
		NGC 4526 & -1.9 & 1-0$^{27}$ & $673.8\pm30.1$ & $79\pm\phantom{1}3^{31}\,$(M) & $8.65\pm0.29^{31}$ & CO\\
		NGC 4593 & \phantom{-}3.0 & 1-0$^{16}$ & $368.1\pm34.4$ & $35\pm\phantom{1}6^{2\phantom{1}}\,$(B) & $6.86\pm0.21^{32}$ & reverberation\\
		NGC 5548 & \phantom{-}0.4 & 1-0$^{16}$ & $212.8\pm31.0$ & $41\pm\phantom{1}6^{2\phantom{1}}\,$(B) & $7.70\pm0.13^{33}$ & reverberation\\
		NGC 7052 & -4.9 & 1-0$^{9\phantom{1}}$ & $683.7\pm81.5$ & $75\pm\phantom{1}1^{34}\,$(M) & $9.41\pm0.05^{34}$ & CO\\
		UGC 3789 & \phantom{-}1.6 & 1-0$^{35}$ & $271.1\pm33.1$ & $43\pm\phantom{1}5^{2\phantom{1}}\,$(B) & $6.99\pm0.09^{20}$ & masers\\
		\hline
	\end{tabular}
	\parbox{0.95\textwidth}{\textbf{Notes:} Column 1 lists the name of each galaxy contained in the final sample of spatially-unresolved galaxies. The morphological classification on the numerical Hubble scale from \href{http://leda.univ-lyon1.fr}{HyperLEDA} is listed in Column 2. Column 3 indicates the CO transition observed. Column 4 lists the (line-of-sight projected) line width and associated uncertainty measured from a Gaussian double peak line profile fit to this spectrum. Column 5 lists the inclination of the CO disc and in parentheses the method used to measure it (D~-~dust morphology, M~-~molecular gas morphology/kinematics, B~-~\textit{B}-band apparent flattening, K~-~other kinematics). Column 6 lists the dynamically-measured SMBH mass using the tracer (or via a Virial estimate for reverberation mapping) listed in Column 7. Column 8 contains other notes about certain galaxies. Footnotes in column 3 indicate the source of the CO observations, and those in columns 6 and 7 the source of the measurement, as follows.
	\textbf{References:} (1)~\cite{Evans+2005ApJS159.197}, (2)~\href{http://leda.univ-lyon1.fr}{HyperLEDA}, (3)~\cite{Kollatschny+2014AandA566.106}, (4)~this work, IRAM project 191-18, (5)~this work, from a \textit{HST} ACS/HRC F550M image, (6)~\cite{Doroshenko+2008ARep52.442}, (7)~\cite{Bertram+2007AandA470.571}, (8)~\cite{Peterson+2004ApJ613.682}, (9)~\cite{OcanaFlaquer+2010AandA518.A9}, (10)~\cite{North+2019MNRAS490.319},  (11)~\cite{Young+2011MNRAS414.940}, (12)~\cite{Cappellari+2006MNRAS366.1126}, (13)~\cite{Smith+2019MNRAS485.4359}, (14)~this work, from a \textit{HST} WFPC2 F814W image, (15)~\cite{Beifiori+2012MNRAS419.2497}, (16)~\cite{Maiolino+1997ApJ485.552}, (17)~\cite{Lodato+2003AandA398.517},  (18)~\cite{Davis+2016MNRAS455.214}, (19)~\cite{Heckman+1989ApJ342.735}, (20)~\cite{Kuo+2011ApJ727.20}, (21)~\cite{Welch+2003ApJ584.260}, (22)~\cite{Cappellari+2013MNRAS432.1709}, (23)~\cite{Schulze+2011ApJ729.21}, (24)~\cite{Onishi+2017MNRAS468.4663}, (25)~this work, from a \textit{HST} WFPC2 F606W image, (26)~\cite{Davis+2018MNRAS473.3818}, (27)~\cite{Combes+2007MNRAS377.1795}, (28)~\cite{Sarzi+2001ApJ550.65}, (29)~\cite{Ford+1994ApJL435.27}, (30)~\cite{Walsh+2013ApJ770.86}, (31)~\cite{Davis+2013Nature494.328},  (32)~\cite{Barth+2013ApJ769.128}, (33)~\cite{Kovacevic+2014ASR54.1414}, (34)~\cite{Smith+2020MNRASsubmitted2} and (35)~this work, OSO 20-m project 2018-04a.} 
\end{table*}

Since we observe the line-of-sight projection of each line width, rather than the intrinsic width, we de-project $W_{50}$ determined by the fit by $\sin i$. Uncertainties in the inclination, which can be significant, are propagated by Monte Carlo sampling as before. The inclinations are determined either from fits to dust features or the resolved CO discs in the literature, ellipse fits to dust discs in archival \textit{Hubble Space Telescope} (\textit{HST}) images, or, where the other methods are not possible, by using the apparent flattenings of the $25\,$mag$\,$arcsec$^{-2}$ $B$-band isophotes and assuming morphology-dependent intrinsic thicknesses as given in HyperLEDA\footnote{The formula for this inclination and for the assumed intrinsic thickness is given online at \href{https://leda.univ-lyon1.fr/leda/param/incl.html}{https://leda.univ-lyon1.fr/leda/param/incl.html}}. \cite{Davis+2011MNRAS414.968} discuss the relative merits (and dangers) of using these methods to infer the inclination of a molecular gas disc. Where the \textit{B}-band isophotes are used, the uncertainties in both the apparent flattenings and the morphological T-types listed in HyperLEDA are propagated into the inclination uncertainties by Monte Carlo sampling. This approach does not work for the face-on galaxy NGC~4388, which appears to be flatter than the inferred intrinsic thickness given its morphological classification, so we adopt a typical inclination uncertainty of $5\degree$. We note this has a negligible effect on the deprojected line width.

\subsection{Correlation fits}
\label{ssec_corrResults}
Given our derived line widths, and the associated SMBH masses and stellar velocity dispersions, we now investigate the correlations between these parameters. We use the \texttt{HYPER-FIT} package \citep{Robotham+2015PASA32.33} via its web interface\footnote{\href{http://hyperfit.icrar.org/}{http://hyperfit.icrar.org/}} to fit both line width to SMBH mass and line width to stellar velocity dispersion. \texttt{HYPER-FIT} seeks to maximise the likelihood function that takes into account the multivariate Gaussian uncertainties on each data point, and allows for the possibility of intrinsic scatter. The use of this approach, in contrast to the traditional forward and/or reverse fits used in many Tully-Fisher relation works \cite[e.g.][]{Tiley+2016MNRAS461.3494}, allows us to include the significant uncertainties on both $M_\mathrm{BH}$ and $W_{50}$, and to minimise the scatter orthogonal to the best-fitting line (rather than only the vertical or horizontal scatter). 

To reduce the covariance between the slope and the intercept, and the error in the intercept, we follow the approach of \cite{Tremaine+2002ApJ574.740}, as is now common practice, and translate the data to bring the median line width closer to zero. We therefore translate the line widths by $2.7\,$dex and fit the general function
\begin{equation}
y = a \left[\log\left(\frac{W_{50}}{\sin i\,\mathrm{km\,s^{-1}}}\right) - 2.7\right] + b\,,
\label{eq_fitFunc}
\end{equation}
where the variable $y$ is an observable quantity - for this work either the SMBH mass or the stellar velocity dispersion. We also determine the total scatter as the root-mean-square deviation (along the $y$-axis) of the data from the best-fitting relation assuming zero measurement errors. Anticipating that the principal application for these relations will be estimating $M_\mathrm{BH}$ from a measured $\Delta V_\mathrm{CO}$, we use the projection of the intrinsic scatter onto the $y$-axis to quantify the tightness of each fit.

We omit from the fits a small number of galaxies that, although having well-constrained SMBH masses and sufficiently double-horned line profiles to yield a robust measurement of $W_{50}$, are nevertheless not considered sufficiently reliable to use. These are indicated in Tables \ref{tab_resolvedData} and \ref{tab_unresolvedData} and are discussed in detail in Section \ref{ssec_outliers}.

Table \ref{tab_correlationResults} lists the results of our fits for both relations (discussed in Sections \ref{ssec_resultVSMBH} and \ref{ssec_resultVSigma}) and for two-different morphologically-selected sub-samples (ETGs and LTGs), in addition to our spatially-resolved and unresolved galaxy samples (and all data/galaxies taken together; see Section \ref{ssec_resUnresCompare}). Seven galaxies are included in both the resolved and unresolved samples, as there are both interferometric and single-dish observations available. The fits for `all' galaxies use the line width with the smaller uncertainty only, almost always from the resolved observations. 

In the following two subsections, we present our results for each correlation and evidence for a morphological dependence. In Section \ref{sec_discussion}, we describe the impacts of using spatially-resolved or unresolved data, compare our results with other host property correlations, and explore the implications of the $\Delta V_\mathrm{CO}{-}M_\mathrm{BH}$ correlation.

\begin{table*}
\begin{center}
\caption{Best-fitting correlations, based on \texttt{HYPER-FIT} fits of Equation \ref{eq_fitFunc}.}
\label{tab_correlationResults}
\begin{tabular}{lccccc}
    \hline
    Dataset & Count & $a$ & $b$ & Total &Intrinsic\\ 
                 &            &    &    & scatter & scatter\\ 
    (1) & (2) & (3) & (4) & (5) & (6)\\
    \hline
    \multicolumn{5}{l}{SMBH mass ($y\equiv \log[M_\mathrm{BH}/\mathrm{M_\odot}]$; Figure \ref{fig_VcSMBH}):}\\
    \hline
    Resolved data  & 25 & $\phantom{1}8.5\pm0.9$ & $7.5\pm0.1$ & 0.6 & $0.5\pm0.1$\\
    Unresolved data & 21 & $10.5\pm2.3$ & $7.6\pm0.2$ & 0.9 & $0.8\pm0.2$\\
    All data & 39 & $\phantom{1}9.2\pm1.1$ & $7.6\pm0.1$ & 0.7 & $0.7\pm0.1$\\

    Resolved ETGs & 16 & $\phantom{1}8.7\pm1.7$ & $7.4\pm0.3$ & 0.6 & $0.6\pm0.2$\\
    Unresolved ETGs & 12 & $12.7\pm4.1$ & $7.3\pm0.5$ & 0.8& $0.7\pm0.3$\\
    All ETGs & 21 & $10.0\pm2.1$ & $7.4\pm0.3$ & 0.8 & $0.7\pm0.2$\\

    Resolved LTGs & \phantom{1}9 & $10.0\pm3.5$ & $7.7\pm0.2$ & 0.5 & $0.4\pm0.1$\\
    Unresolved LTGs & \phantom{1}9 & $12.0\pm9.4$ & $7.8\pm0.5$ & 1.3 & $1.3\pm1.0$\\
    All LTGs & 18 & $10.2\pm2.8$ & $7.7\pm0.3$ & 0.8 & $0.8\pm0.3$\\
    \\
    \multicolumn{5}{l}{Stellar velocity dispersion ($y\equiv\log[\sigma_\ast/\mathrm{km\,s^{-1}}]$; Figure \ref{fig_VcSigma}):}\\
    \hline
    Resolved data & 25 & $1.1\pm0.1$ & $2.20\pm0.02$ & 0.09 & $0.09\pm0.02$\\
    Unresolved data & 21 & $0.9\pm0.2$ & $2.22\pm0.02$ & 0.09 & $0.08\pm0.02$\\
    All data & 39 & $1.1\pm0.1$ & $2.21\pm0.02$ & 0.10 & $0.10\pm0.01$\\
    
    Resolved ETGs & 16 & $1.0\pm0.2$ & $2.22\pm0.03$ & 0.08 & $0.07\pm0.02$\\
    Unresolved ETGs & 12 & $0.9\pm0.3$ & $2.22\pm0.04$ & 0.06 & $0.06\pm0.02$ \\
    All ETGs & 21 & $1.0\pm0.2$ & $2.23\pm0.03$ & 0.08 & $0.07\pm0.01$\\
    
    Resolved LTGs & \phantom{1}9 & $1.0\pm0.5$ & $2.17\pm0.05$ & 0.11 & $0.10\pm0.04$\\
    Unresolved LTGs & \phantom{1}9 & $0.8\pm0.6$ & $2.21\pm0.04$ & 0.11 & $0.11\pm0.04$\\
    All LTGs & 18 & $1.1\pm0.4$ & $2.20\pm0.04$ & 0.12 & $0.11\pm0.03$ \\
    \hline
\end{tabular}
\end{center}
\parbox{0.95\textwidth}{\textbf{Notes:} Column 1 lists each sample of galaxies, Column 2 the number of galaxies in that sample. Columns 3 and 4 list the parameters $a$ and $b$, respectively, measured by fitting Equation \protect\ref{eq_fitFunc} to the data of that sample with \texttt{HYPER-FIT}. Column 5 lists the total scatter, defined as the root-mean-square deviation along the $y$-axis between the data and the best-fitting relation, of that sample. Column 6 lists the intrinsic scatter, projected along the $y$-axis, of that sample.}
\end{table*}

\subsection{The CO $\Delta V_\mathrm{CO}{-}M_\mathrm{BH}$ correlation}
\label{ssec_resultVSMBH}
\begin{figure}
	\includegraphics[trim={0.5cm 0.75cm 1.5cm 1cm}, width=0.5\textwidth]{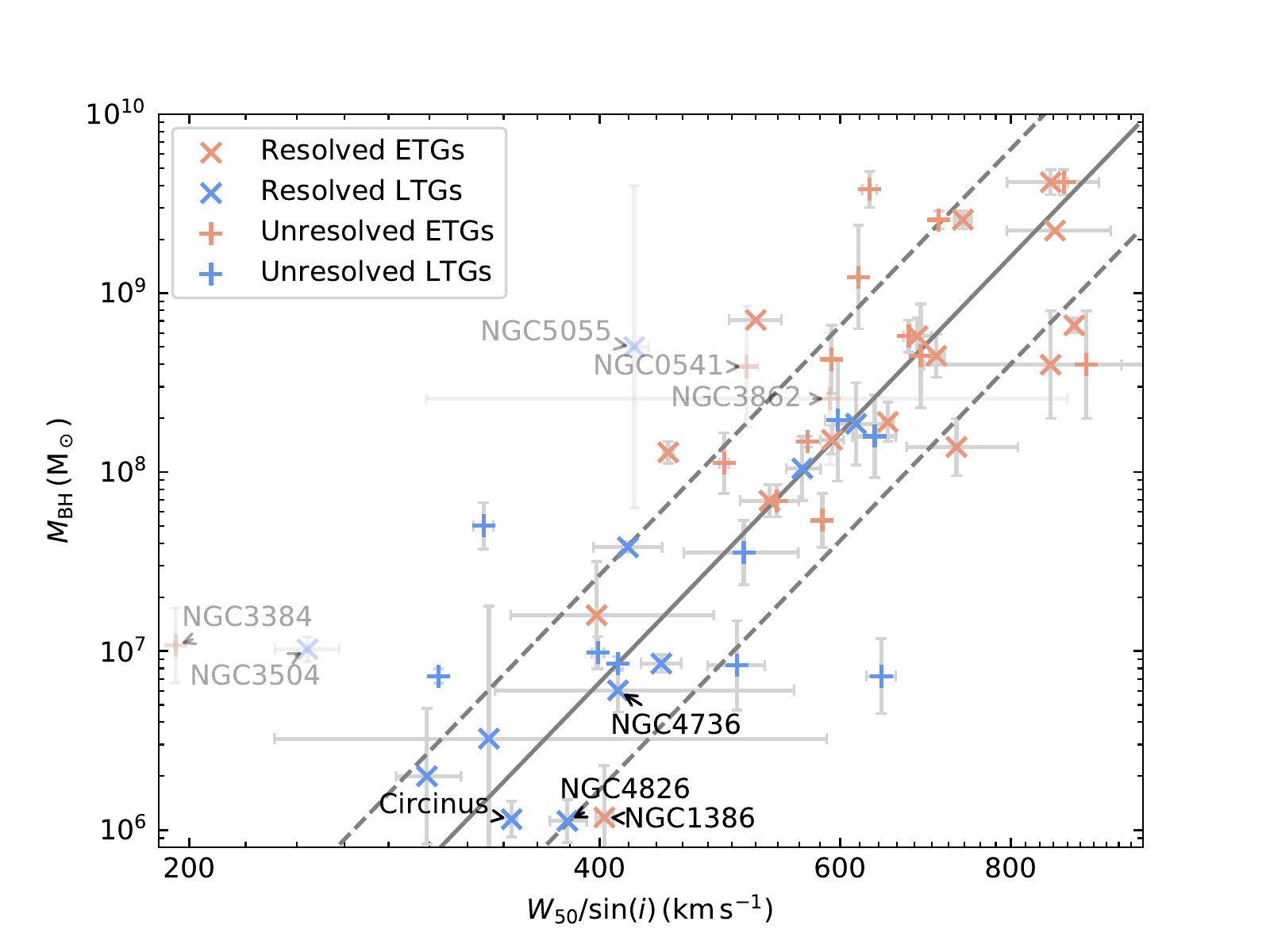}
    \caption{Correlation between deprojected line width ($W_{50}/\sin i$) and SMBH mass ($M_\mathrm{BH}$) for our sample galaxies. Colours indicate whether a galaxy is classified as early- (red) or late-type (blue), while the markers indicate whether it belongs to our spatially-resolved ($\times$) or unresolved (+) sample. Some galaxies appear in both samples, and hence appear twice on this plot at the same $M_\mathrm{BH}$. Labelled galaxies are discussed in Section \ref{ssec_outliers}. The tightest correlation determined from the resolved sample is indicated by the dark grey solid line, with the $1\sigma$ intrinsic scatter indicated by the dashed dark grey lines. Error bars are shown in pale grey. Faded galaxies are excluded from the correlation fits.}
    \label{fig_VcSMBH}
\end{figure}

Figure \ref{fig_VcSMBH} shows the $\Delta V_\mathrm{CO}{-}M_\mathrm{BH}$ correlation of all our data and the best-fitting relation derived using only the galaxies with spatially-resolved observations. This is the tightest correlation we find, with
\begin{equation}
\log\left(\frac{M_\mathrm{BH}}{\mathrm{M_\odot}}\right) = (8.5\pm0.9)\left[\log\left(\frac{W_{50}}{\sin i\,\mathrm{km\,s^{-1}}}\right) - 2.7\right] + (7.5\pm0.1)\,,
\end{equation}
with a total scatter in the $\log M_\mathrm{BH}$ direction of $0.6\,$dex, dominated by the intrinsic scatter of $0.5\,$dex.

The most significant systematic deviations from this fit are when we restrict the sample to LTGs. When we consider only the spatially-resolved LTGs, the slope steepens to $10.0\pm3.5$, in agreement with the results of \cite{Davis+2019ApJ877.64} from \ion{H}{I} and invoking the disc-halo conspiracy. The sample of unresolved LTGs do not adequately constrain the relation's slope,  and exhibit a much higher total (and intrinsic) scatter. We discuss this deviation further in Section \ref{ssec_resUnresCompare}.

\subsection{The $\Delta V_\mathrm{CO}-\sigma_\ast$ correlation}
\label{ssec_resultVSigma}
\begin{figure}
	\includegraphics[trim={0.5cm 1cm 1.5cm 1cm}, width=\columnwidth]{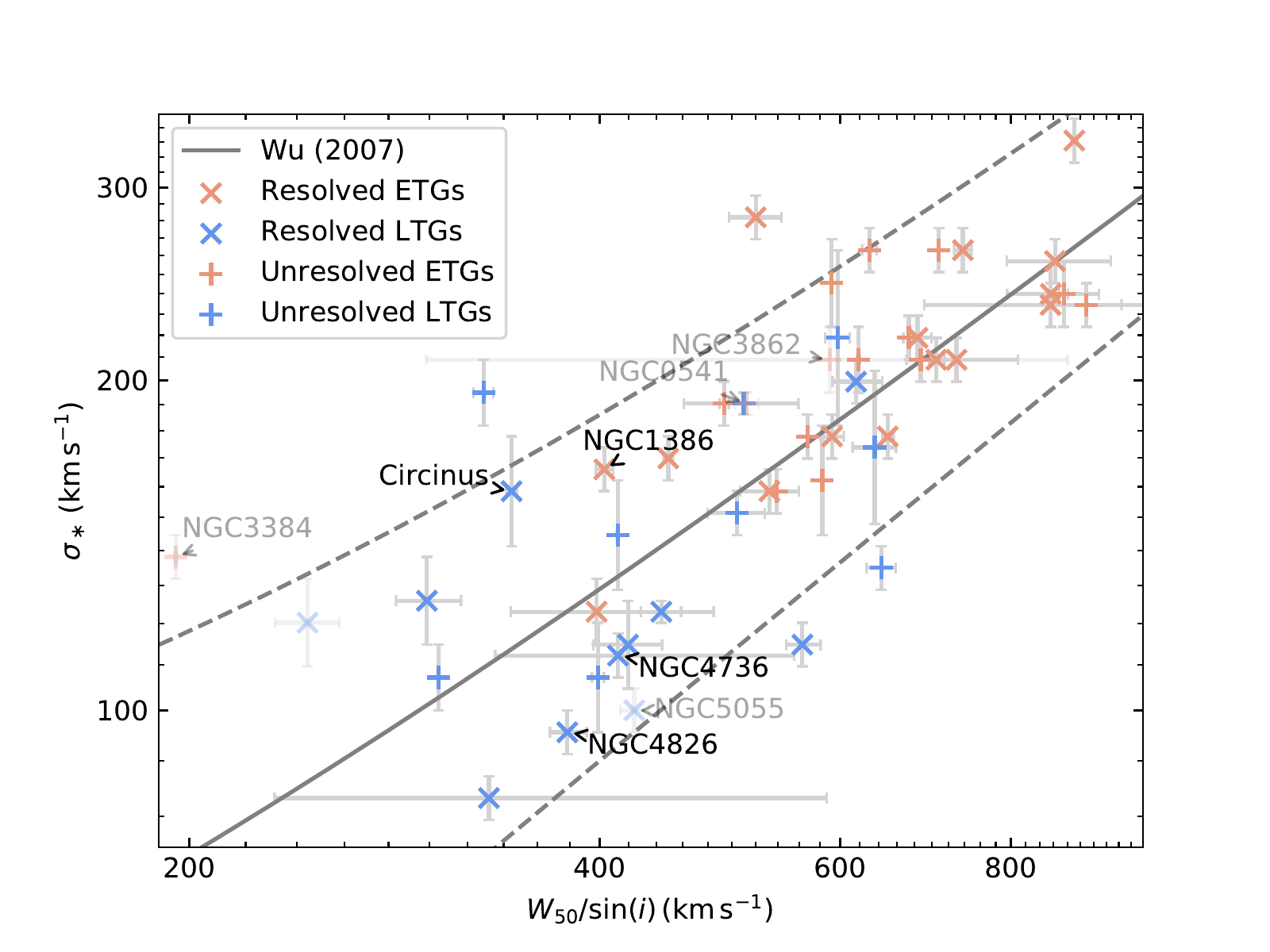}
    \caption{Correlation between deprojected line width ($W_{50}/\sin i$) and stellar velocity dispersion ($\sigma_\ast$) for our sample galaxies. Colours indicate whether a galaxy is classified as early- (red) or late-type (blue), while the markers indicate whether it belongs to our spatially-resolved ($\times$) or unresolved (+) sample. Some galaxies appear in both samples, and hence appear twice on this plot at the same $\sigma_\ast$. Labelled galaxies are discussed in Section \ref{ssec_outliers}. The best-fitting correlation found by \protect\cite{Wu2007ApJ657.177} amongst local Seyfert galaxies is indicated by the solid dark grey line, with the $1\sigma$ intrinsic scatter indicated by the dashed dark grey lines. Error bars are shown in pale grey. Faded galaxies are excluded from the correlation fits.}
    \label{fig_VcSigma}
\end{figure}

Figure \ref{fig_VcSigma} shows the $\Delta V_\mathrm{CO}{-}\sigma_\ast$ correlation using our data and the stellar velocity dispersions compiled by \cite{vdBosch2016ApJ831.134}. These velocity dispersions are those available in the literature that most closely approximate the dispersion within $1\,R_\mathrm{e}$.

We find that the dispersions are consistent with a linear relationship between CO line width and $\sigma_\ast$ for all sub-samples. The best-fitting relation, from the sample of spatially-resolved observations of galaxies of all morphologies, is

\begin{equation}
\log\left(\frac{\sigma_\ast}{\mathrm{km\,s^{-1}}}\right) = (1.1\pm0.1)\left[\log\left(\frac{W_{50}}{\sin i\,\mathrm{km\,s^{-1}}}\right) - 2.7\right] + (2.20\pm0.02)\,.
\end{equation}
There is a systematic trend of the intrinsic scatter with morphology, ETGs having an intrinsic scatter of $0.07\pm0.02\,$dex in the $\log\sigma_\ast$ direction, whereas LTGs have $0.10\pm0.04\,$dex.

The most recent works investigating $\Delta V_\ion{H}{I}{-}\sigma_\ast$ have also found morphologically-varying results. In ETGs, \cite{Serra+2016MNRAS460.1382} found a linear relation with total scatter of $12\%$, whereas in late-type spirals \cite{Davis+2019ApJ877.64} excluded a linear relation, obtaining $\sigma_\ast\propto\Delta V_\ion{H}{I}^{1.55\pm0.25}$. This accords with the early work of \cite{Ho2007ApJ668.94}, that indicated that $\Delta V/\sigma_\ast$ varies systematically with Hubble-type, albeit with less compelling data. The literature on the $\Delta V_\mathrm{CO}{-}\sigma_\ast$ has not considered a potential morphological variation systematically.

We, however, find no significant deviation from a linear relation for either early- or late-type galaxies. This is perhaps surprising, as using the disc-halo conspiracy for LTGs to equate $\Delta V_\mathrm{CO}=\Delta V_\ion{H}{I}$, the results of \cite{Davis+2019ApJ877.64} would predict otherwise. The small intrinsic scatter for ETGs agrees with the results from JAM modelling of \cite{Cappellari+2013MNRAS432.1709}, that indicate a tight correlation between the rotation curve at these scales and $\sigma_\ast$.

\section{Discussion}
\label{sec_discussion}
In this section we first discuss the benefits of using spatially-resolved rather than unresolved observations. We then evaluate the biases associated with our sample, account for the galaxies excluded in our fits, and finally discuss the utility of our correlations for estimating SMBH masses. 

\subsection{Spatially-resolved vs unresolved observations}
\label{ssec_resUnresCompare}
In Section \ref{sec_data} we outlined the advantages of using spatially-resolved observations of the CO emission instead of single-dish observations. These advantages are clear in our results. 

First, the spatially-resolved observations have much higher sensitivities. Since the uncertainties on our line width measurements are derived from Monte Carlo fits to simulated data with noise characteristic of the real spectra, these improved sensitivities lead to smaller line width uncertainties (of order $1-10\,$km$\,$s$^{-1}$ rather than $10-50\,$km$\,$s$^{-1}$), that are also smaller fractions of the channel widths. This in turn leads to best-fitting correlations with systematically smaller uncertainties. While present, the effect on the total scatters (dominated by the intrinsic scatters) is less significant.

Second, spatially-resolved kinematics enable improved sample selection, ensuring that galaxies with disturbed kinematics can be omitted more robustly. This reduces the measured intrinsic scatters in our $\Delta V_\mathrm{CO}{-}M_\mathrm{BH}$ and $\Delta V_\mathrm{CO}{-}\sigma_\ast$ correlations when only resolved observations are used. 

Third, the relatively small primary beams of large single-dish observations (e.g. $22\arcsec$ for the IRAM 30-m telescope at \mbox{$^{12}$CO(1-0)}) mean that observations may not reach the flat parts of the rotation curves of nearby galaxies. In spatially-resolved observations, not only do we benefit from the larger primary beams of the smaller individual antennae (e.g. $55\arcsec$ for ALMA's 12-m dishes at \mbox{$^{12}$CO(1-0)}), but pointing errors can also be straightforwardly diagnosed. 

We briefly note that, in addition to the primary beam, another spatial scale generally relevant for interferometric observations is the maximum resolvable scale, i.e. the largest spatial structure to which an array configuration is sensitive, set by the shortest baseline of the array. However, CO emission is generally patchy, and thus most of it generally remains detectable even when only an extended array configuration is used. In addition, as the emission of a rotating disc extends only over a small spatial scale in any given channel (typically of the order of the disc minor axis along one direction only), we are likely to recover (most of) it all the way to the flat part of the rotation curve, provided the emission extends that far.

We have also argued that a sharp-edged double-horned line profile arises from an emitting exponential disc that reaches the flat part of the rotation curve. However, such a profile can also occur if the disc is sharply truncated, whether the flat part is reached or not. If such a disc is truncated before the flat part, the measured line width will be strictly narrower than would be found if the disc extended further. Similarly, if the primary beam is too compact to reach the flat part, the measured line width will be artificially narrowed.

In each spatially-resolved sample galaxy, we can easily assess whether the CO emission reaches the flat part of the rotation curve by examining the kinematic major-axis position velocity diagram, and thus straightforwardly determine whether the primary beam is too small to recover all of the emission. For each of our spatially-unresolved sample galaxies, we assess this issue as follows. As low-\textit{J} CO emission generally follows dust, the extent of the dust disc in \textit{HST} optical images can be assumed to be the same as that of the CO emission. Although some of these galaxies have dust extending to larger radii than would be reached by the primary beams of our single-dish telescopes, we have verified that these galaxies do not exhibit systematically narrower line profiles. However, the galaxies with the smallest dust extents ($\lesssim0.02\,R_{25}$, where $R_\mathrm{25}$ is the radius of the $25\,$mag$\,$arcsec$^{-2}$ $B$-band isophote listed in HyperLEDA) are biased to narrower line widths. We therefore exclude them from this sample.

Notwithstanding the concerns described above, we do not find a statistically-significant difference between the correlations determined from the spatially-resolved and unresolved samples, and Figure \ref{fig_VcSMBH} further indicates that the unresolved ETGs follow the relation of all resolved galaxies closely; the significant outliers are all sufficiently explained in Section \ref{ssec_outliers} as resulting from observations of CO emission that may not reach the flat part of the rotation curve. The only substantial discrepancy is found for the $\Delta V_\mathrm{CO}{-}M_\mathrm{BH}$ relation using unresolved LTGs, with a much greater uncertainty in the slope (and to a lesser extent in the zero point) and significantly larger intrinsic scatter than those of all resolved data and ETGs. These galaxies are likely to be preferentially affected by the aforementioned issues, as they are systematically nearer and have more slowly rising rotation curves. It is therefore unsurprising that we find greater uncertainties in their best-fitting parameters and a larger intrinsic scatter. Interestingly, the spatially-resolved LTG sample, for which we can exclude galaxies with disturbed kinematics, actually exhibits a slightly smaller intrinsic scatter.

For all the reasons discussed above, we conclude that spatially-resolved observations are to be preferred when calibrating (and using) the $\Delta V_\mathrm{CO}{-}M_\mathrm{BH}$ correlation. We further suggest that the improvements offered by the use of spatially-resolved observations have wider applicability, particularly when calibrating the CO TFR. Interferometric observations in the ALMA era will therefore allow sample selections significantly more robust than was previously possible, with associated improvements of the accuracy of the slopes, zero-points and intrinsic scatters of the determined relations.


\subsection{Selection biases}
Our sample was selected from galaxies with dynamically-measured SMBH masses in the literature and CO observations, and thus cannot be considered a statistically-representative sample of galaxies. \cite{Shankar+2016MNRAS460.3119} discussed the biased population of galaxies with dynamical SMBH mass measurements, due to the need to resolve the scales on which the SMBH dominates the potential. Additionally, the SMBH mass measurements and CO observations are highly heterogenous, being derived using different dynamical tracers and resolving different physical scales for the former, and with different primary beams, spectral resolutions and sensitivities for the latter.

The heterogenous CO observations have all been homogenised following the prescriptions discussed in Sections \ref{ssec_resolvedSample} and \ref{ssec_unresolvedSample}. These procedures do not bias our conclusions regarding the correlations, but the selection of only sharp-edged double-horned profiles arguably limits the applicability of the correlations to similar CO spectra only. 

\begin{figure}
	\includegraphics[trim={0.5cm 1cm 1.5cm 1cm}, width=\columnwidth]{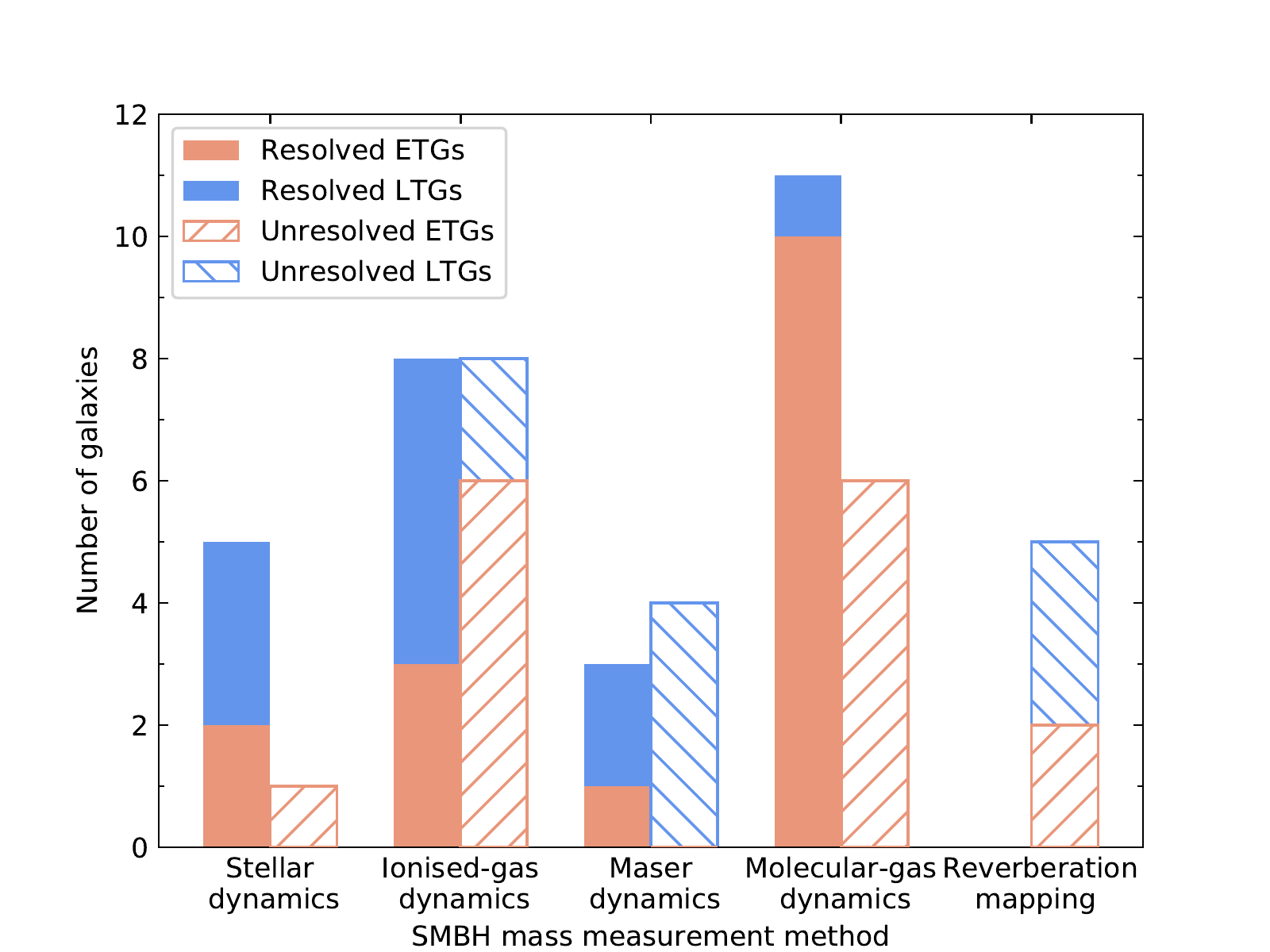}
	\vspace{0.1cm}
    \caption{SMBH mass measurement methods of our sample galaxies. Early-type galaxies are shown in red, late-type galaxies in blue. The spatially-resolved sample is shown in solid colour, the unresolved sample as hatched areas. Galaxies appear in both resolved and unresolved samples for a given method where both interferometric and single-dish observations exist.}
    \label{fig_methods}
\end{figure}

The SMBH masses used in this paper are drawn from the large variety of measurements available in the literature. Due to the differing selection criteria, very few SMBH mass measurements have been cross-checked with multiple tracers and/or methods, and those that have suggest mass measurements can vary by factors of 2--4 \citep[e.g.][]{Kormendy+2013ARAA51.511}. Figure \ref{fig_methods} shows the variety of mass measurement methods used. Maser dynamics are, in principle, the most precise method as masers probe the spatial scales closest to the SMBHs, while the uncertainty in the scaling factor used in reverberation mapping ($f$; meant to account for the broad-line region geometry and line-of-sight velocity dispersion anisotropy) implies these measurements are generally the least reliable \citep[e.g.][]{Pancoast+2014MNRAS445.3055, MejiaRestrepo+2018NatAs2.63, Campitiello+2019arXiv190700986}. The associated SMBH mass uncertainties are taken into account by the \texttt{HYPER-FIT} routine, so that lower-quality measurements do not bias our results. Although it would be preferable to use homogeneously-measured SMBH masses, to impose this requirement would excessively reduce the sample size.

The $M_\mathrm{BH}{-}\sigma_\ast$ relation is the tightest known correlation between SMBH mass and host property, and it is customarily interpreted as indicating the primary co-evolutionary path for SMBH growth. Galaxies that are outliers on the $M_\mathrm{BH}{-}\sigma_\ast$ plane are therefore likely to have had unusual evolutionary pathways, and it is unlikely they will follow other host--SMBH correlations. Since we argued in Section \ref{ssec_COlinesTraceBaryons} that the CO line width of a galaxy traces the same baryonic matter as $\sigma_\ast$, any galaxy that genuinely deviates from the $M_\mathrm{BH}{-}\sigma_\ast$ relation is also likely to deviate from the $\Delta V_\mathrm{CO}{-}M_\mathrm{BH}$ relation. Galaxies consistent with only one of these relations may be genuine outliers, or may indicate one of the quantities has been incorrectly measured.

\begin{figure}
	\includegraphics[trim={0.5cm 1cm 1.5cm 1cm}, width=\columnwidth]{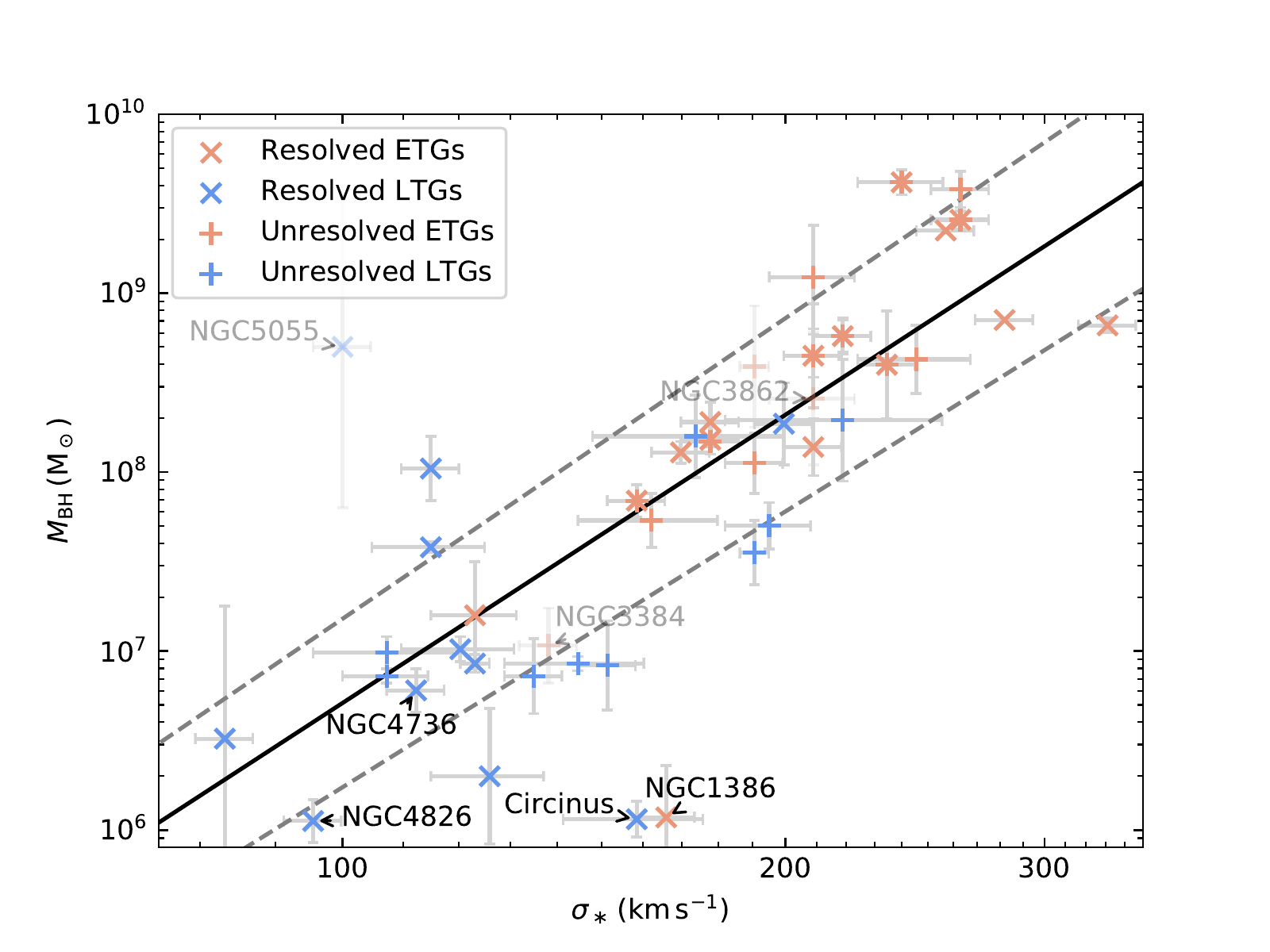}
    \caption{Correlation between stellar velocity dispersion ($\sigma_\ast$) and SMBH mass ($M_\mathrm{BH}$) for our sample galaxies. Colours indicate whether a galaxy is classified as early- (red) or late-type (blue), while the markers indicate whether it belongs to our spatially-resolved ($\times$) or unresolved (+) sample. Some galaxies appear in both samples, and hence are indicated with both symbols superimposed. The $M_\mathrm{BH}{-}\sigma_\ast$ relation found by \protect\cite{vdBosch2016ApJ831.134} is indicated by the solid black line, with the $1\sigma$ intrinsic scatter indicated by the dashed dark grey lines. Labelled galaxies are discussed in Section \ref{ssec_outliers}. Error bars are shown in pale grey. Faded galaxies are excluded from the correlation fits.}
   \label{fig_mSigma}
\end{figure}

Figure \ref{fig_mSigma} shows the $M_\mathrm{BH}{-}\sigma_\ast$ relation for our sample galaxies. A few galaxies are clearly outliers and are discussed in Section \ref{ssec_outliers}. The remainder of the sample galaxies all follow the empirical $M_\mathrm{BH}{-}\sigma_\ast$ relation and are not significantly biased in their distribution in the $M_\mathrm{BH}{-}\sigma_\ast$ plane. We do not, however, sample the significant population of galaxies with $\sigma_\ast<100\,$km$\,$s$^{-1}$, for which very few SMBH masses are available, and that appear to deviate from the $M_\mathrm{BH}{-}\sigma_\ast$ relation \citep[e.g.][]{vdBosch2016ApJ831.134}.

\begin{figure}
	\includegraphics[trim={0.5cm 1cm 1.5cm 1cm}, width=\columnwidth]{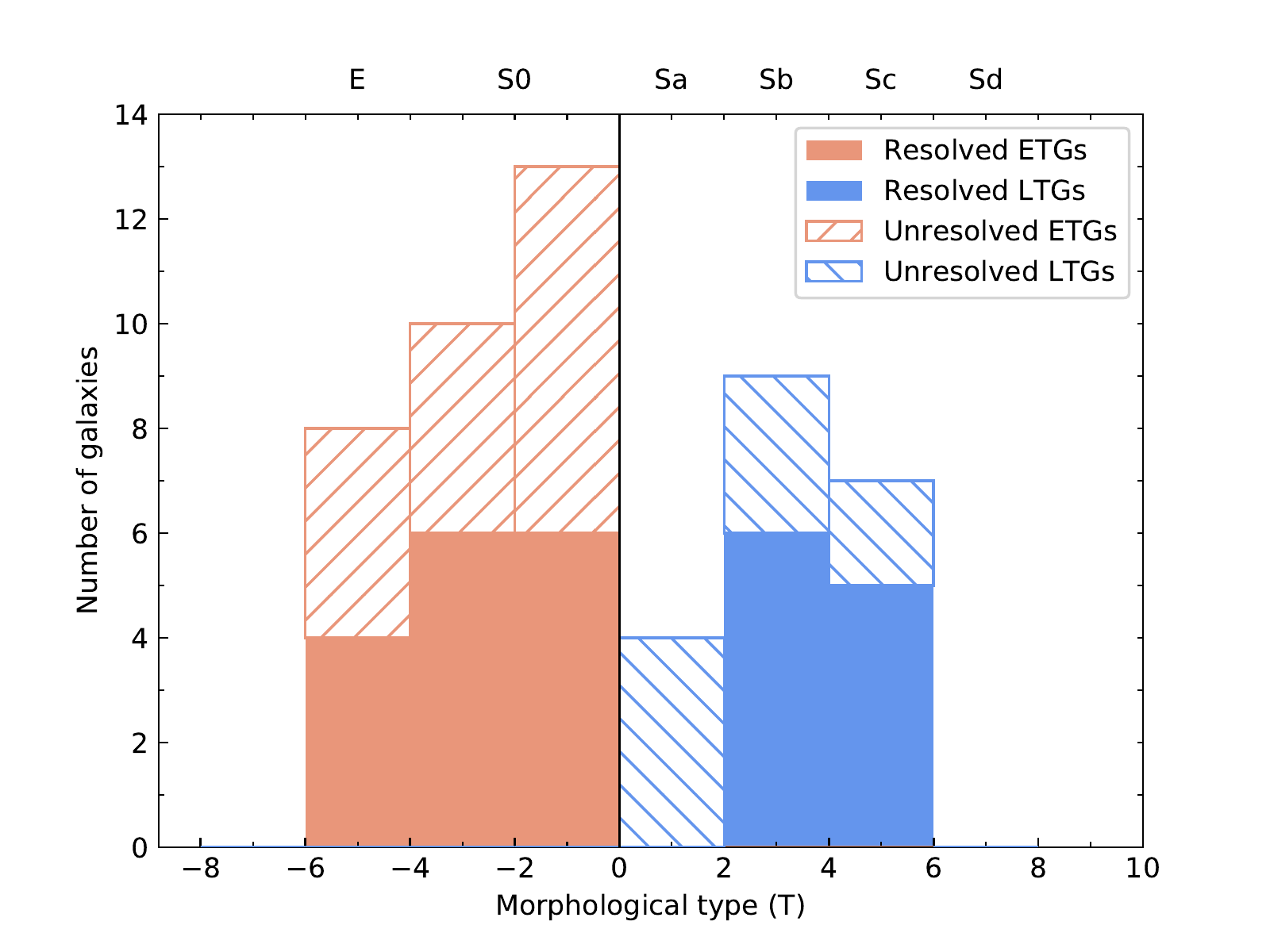}
    \caption{Morphological type distribution of our sample galaxies, according to the numerical Hubble type listed in HyperLEDA, with $T\leq0$ classified as early type and $T>0$ as late type.} 
    \label{fig_morphDist}
\end{figure}

Our samples are necessarily biased towards CO-bright galaxies. The CO content of a galaxy varies with morphology, the latest representative surveys finding that only about $20{-}30\%$ of ETGs host detectable molecular gas reservoirs \citep{Combes+2007MNRAS377.1795, Young+2011MNRAS414.940}. However, this detection rate is independent of mass, size and environment, suggesting that CO-rich ETGs are `normal' ETGs \citep{Davis+2019MNRAS486.1404}. CO emission is also detected in ${\approx}85\%$ of LTGs \cite[][]{Young+1995ApJS98.219, Saintonge+2017ApJS233.22}. As a result, our samples encompass a wide range of morphologies, as shown in Figure \ref{fig_morphDist}. 

In Sections \ref{ssec_resultVSMBH} and \ref{ssec_resultVSigma} we discussed whether our results depend on galaxy morphology. We adopted the HyperLEDA morphological classifications on the numerical Hubble scale, and classified galaxies with T-types less than or equal to 0 as ETGs and the others as LTGs. The morphological type distribution of our sample is illustrated in Figure \ref{fig_morphDist} by colour and the solid vertical line.

Our samples are thus necessarily limited by the existing biases in the SMBH masses that have been measured, and to galaxies with CO emission. Nevertheless, we do not find evidence of a systematic deviation in the $\Delta V_\mathrm{CO}{-}M_\mathrm{BH}$ correlation as a function of morphology, but do caution that this issue requires further study going beyond the coarse classification used here.

\subsection{Outliers}
\label{ssec_outliers}
We have excluded a few galaxies from our fits even though they have well-resolved and sharp-edged CO lines and dynamically-measured SMBH masses. Each of these is indicated in Tables \ref{tab_resolvedData} and \ref{tab_unresolvedData} and is discussed below.

Three galaxies in our samples are known to be outliers in the $M_\mathrm{BH}{-}\sigma_\ast$ relation: Circinus, \mbox{NGC 1386} and \mbox{NGC 5055}. The former two galaxies have SMBH measurements too small for their associated velocity dispersions by about an order of magnitude. However, they are consistent with our $\Delta V_\mathrm{CO}{-}M_\mathrm{BH}$ relation, their positions on the $\Delta V_\mathrm{CO}{-}\sigma_\ast$ relation (Figure \ref{fig_VcSigma}) compensating. For Circinus, \cite{Davis+2019ApJ877.64} find a similar behaviour using $\Delta V_\ion{H}{I}$, and suggest its central stellar velocity dispersion may be anomalous. We therefore elect to include both galaxies in our fits. 

\mbox{NGC 5055} has a SMBH mass measurement too large by around two orders of magnitude compared to that predicted from either the $M_\mathrm{BH}{-}\sigma_\ast$ relation or our $\Delta V_\mathrm{CO}{-}M_\mathrm{BH}$ correlation. This mass is based on Fabry-Perot spectroscopy taken at the Canada-France-Hawaii Telescope \citep{BlaisOuellette+2004AandA420.147}, in which the central $300\,$pc exhibits dual velocity components, one consistent with the overall galactic rotation, the other with a counter-rotating disc (from which the SMBH mass is derived) or a bipolar outflow. As this SMBH mass is suspect \citep[see][]{Graham2008PASA25.167}, we exclude this object from our fits. 

We also omit the dwarf galaxy NGC~3504, which is an outlier from the $\Delta V_\mathrm{CO}{-}M_\mathrm{BH}$ relation. As the only dwarf galaxy in the sample it would otherwise have a disproportionate effect on the determined slope and scatter, and dwarfs do not appear to follow the other SMBH--host galaxy property correlations (see e.g.$\,$Figure 1 of \citealt{vdBosch2016ApJ831.134}).

In the unresolved sample, we omit \mbox{NGC 541}, \mbox{NGC 3384} and \mbox{NGC 3862}. \mbox{NGC 541} has the most compact dust disc of the galaxies observed, extending to only $0.01\,R_\mathrm{25}$, and a narrow line width consistent with our discussion in Section \ref{ssec_resUnresCompare}. There is no dust disc visible in \mbox{NGC 3384}, but the CO line width is very narrow. \mbox{NGC 3862} has a dust disc of only $0.02\,R_\mathrm{25}$ that is also nearly face-on, implying that the inclination uncertainties are very large.


\mbox{NGC 4736} and \mbox{NGC 4826} are both outliers from the fundamental plane and were therefore excluded from previous fundamental plane parameter -- SMBH mass correlations \citep{vdBosch2016ApJ831.134}. However, they appear consistent with both correlations investigated in this paper, and with the $M_\mathrm{BH}{-}\sigma_\ast$ relation, so they are included in this work.

Finally \mbox{Mrk 590}, \mbox{NGC 524} and \mbox{NGC 4303} all have inclinations below $30\degree$. Studies of the Tully-Fisher relation exclude galaxies at such low inclinations \cite[e.g.][]{TullyFisher1977AandA54.661, Pierce+1988ApJ330.579, Davis+2011MNRAS414.968, Tiley+2016MNRAS461.3494, Topal+2018MNRAS479.3319}. This is because customary approaches to measuring inclination, such as fitting ellipses to features assumed to be intrinsically circular, respond only weakly to varying inclinations at $i{\approx}0\degree$, but the line-of-sight projected velocity responds strongly ($v_\mathrm{los}{\propto}\sin i$). In our work, we allow these galaxies to remain (and omit to label them in Figures \ref{fig_VcSMBH}, \ref{fig_VcSigma}, and \ref{fig_mSigma}) as the large uncertainties have been propagated through, and thus these galaxies have low statistical weights. 

\subsection{Comparison with other correlations}
\label{ssec_litCompare}
The best-fitting correlations for our samples (Table \ref{tab_correlationResults}) can be compared to the extensive literature on other correlations. Correlations between $\Delta V_\ion{H}{I}$ at halo scales and either $M_\mathrm{BH}$ or $\sigma_\ast$ have already been discussed in Sections \ref{ssec_resultVSMBH} and \ref{ssec_resultVSigma}, so we only briefly summarise these findings here before considering other correlations.

We find a close correlation between $\Delta V_\mathrm{CO}$ and $M_\mathrm{BH}$, as would be expected from the simple arguments presented in Section \ref{ssec_COlinesTraceBaryons}. We find a steeper slope for LTGs than for ETGs; the LTG result is in agreement with the slope and intrinsic scatter found using \ion{H}{I} line widths \citep{Davis+2019ApJ877.64}. When computing a single relation for galaxies of all morphologies, we obtain an intrinsic scatter of $0.5\pm0.1\,$dex, identical to that found using \ion{H}{I} in the equivalent sample of \cite{Beifiori+2012MNRAS419.2497}.
 
We find a linear $\Delta V_\mathrm{CO}{-}\sigma_\ast$ correlation regardless of galaxy morphology.  For ETGs, this agrees with the \ion{H}{I} results of \cite{Serra+2016MNRAS460.1382}, even though we cannot assume a baryon-halo conspiracy holds. However, contrary to us, \cite{Davis+2019ApJ877.64} exclude a linear relation in LTGs. \cite{Wu2007ApJ657.177} fit a linear relation to local Seyferts, leading to larger uncertainties in the coefficients, but did not estimate the scatter. We fit the line widths measured by \cite{Wu2007ApJ657.177}, that are also FWHM but are not based on profile fits, with Equation \ref{eq_fitFunc}, yielding the best-fitting parameters $a{=}0.71\pm0.12$ and $b{=}2.21\pm0.02$, with a total scatter of $2.1\,$dex, of which the intrinsic scatter is $0.13\pm0.02\,$dex. The total scatter is dominated by the large uncertainties of the sample stellar velocity dispersions. Our intrinsic scatters are similarly small, although we do not have a Seyfert galaxy sample to directly compare.

The tightest correlations between SMBH masses and host properties have intrinsic scatters comparable to ours. The careful analysis of  \cite{Kormendy+2013ARAA51.511} showed that for classical bulges and ellipticals, the $M_\mathrm{BH}{-}\sigma_\ast$ relation has an intrinsic scatter of $0.29\,$dex. Using a less rigorously selected and larger sample of galaxies, \cite{Beifiori+2012MNRAS419.2497} determined a total scatter of $0.41\pm0.06\,$dex, dominated by the intrinsic scatter of $0.36\pm0.07\,$dex. Our tightest correlation is for spatially-resolved galaxies, with a total scatter of $0.6\,$dex, that is also dominated by the intrinsic scatter of $0.5\,$dex. 

The other major bulge correlations are those with bulge luminosity and mass. \cite{Beifiori+2012MNRAS419.2497} determined intrinsic scatters of $0.58\pm0.11\,$dex and $0.46\pm0.07\,$dex, respectively, while \cite{Kormendy+2013ARAA51.511} found $0.3\,$dex and $0.28\,$dex, respectively. Our results are thus comparably tight to the bulge correlations.

Looser $M_\mathrm{BH}$--host property correlations, including those with S\'ersic index \citep{Graham+2007ApJ655.77, Davis+2017MNRAS471.2187}, spiral arm pitch angle \citep{Seigar+2008ApJL678.93, Davis+2017MNRAS471.2187} and total galaxy light or mass \citep{Jahnke+2009ApJL706.215, Bennert+2010ApJ708.1507, Merloni+2010ApJ708.137, Davis+2018ApJ869.113}, have also been proposed and investigated over the last two decades. \cite{Beifiori+2012MNRAS419.2497} investigated several of these, finding typical intrinsic scatters of $0.5-0.6\,$ dex, similar to or slightly larger than our results.

We therefore conclude that the correlations we have probed are at least no worse than many of the correlations well-established in the literature. However, our $M_\mathrm{BH}{-}\Delta V_\mathrm{CO}$ correlation is not so tight as to outperform those commonly used to estimate SMBH masses. In the absence of a measurement of $\sigma_\ast$, and where a bulge decomposition is either too difficult or too laborious, use of the CO line width is thus a competitive estimator of a galaxy's SMBH mass.

\subsection{Utility for estimating SMBH masses}
\label{ssec_application}
To illustrate the use of CO line widths as SMBH mass estimators, we construct a SMBH mass function (see review by \citealt{Kelly+2012AdAst2012E.7}) from the \cite{Tiley+2016MNRAS461.3494} sample of 207 CO(1-0) line widths, that were measured in a manner identical to that in this work. These observations were originally obtained as part of the COLD GASS survey \citep{Saintonge+2011MNRAS415.32}, with the IRAM 30-m telescope. The sample is purely mass-selected to be representative of galaxies in the local universe with $\log\left( M_\ast/\mathrm{M_\odot}\right) > 10$, that corresponds to $M_\mathrm{BH}\gtrsim10^{6.8}\,\mathrm{M_\odot}$ using the correlation of \cite{Beifiori+2012MNRAS419.2497}. The \cite{Tiley+2016MNRAS461.3494} sample also contains some galaxies below $10^{10}\,$M$_\odot$, that were later published in the extended COLD GASS survey (xCOLDGASS; \citealt{Saintonge+2017ApJS233.22}), but are not necessarily statistically-representative of these galaxies. Removing the galaxies with stellar masses less than $10^{10}\,$M$_\odot$ makes only a marginal change to the derived SMBH mass function (and this only at $M_\mathrm{BH}<10^{6.8}\,$M$_\odot$), even though it makes a substantial change to the expected distribution of galaxies in a volume-limited sample.

\begin{figure}
	\includegraphics[trim={0.5cm 1cm 1.5cm 1cm}, width=0.43\textwidth]{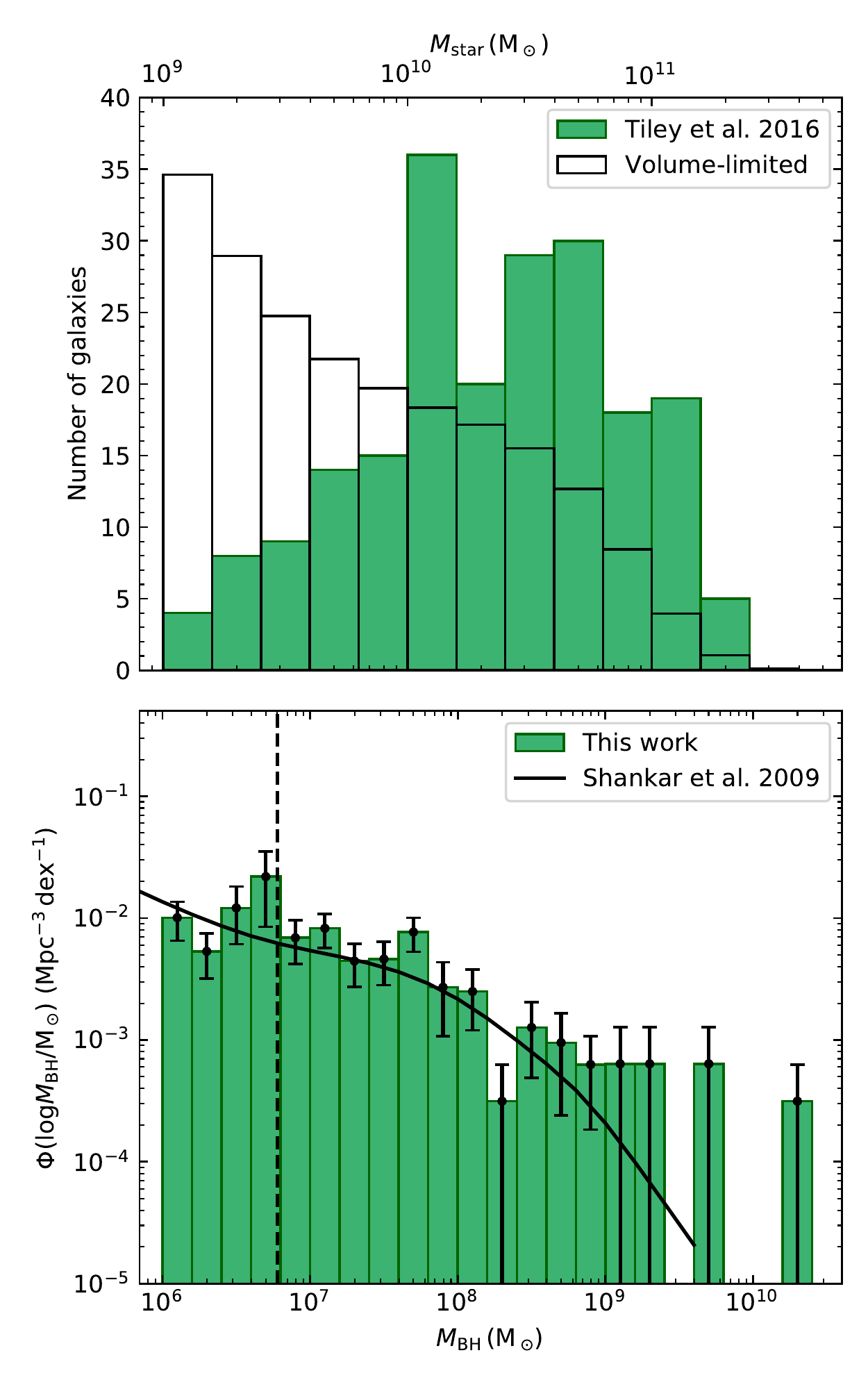}
    \caption{\textbf{Top panel:} Galaxy stellar mass function of the \protect\cite{Tiley+2016MNRAS461.3494} sample drawn from COLD GASS (green histogram), and of a purely volume-limited sample of equal size following the galaxy stellar mass function of \protect\cite{Baldry+2012MNRAS421.621} (black histogram). \textbf{Bottom panel:} Local SMBH mass function derived from the \protect\cite{Tiley+2016MNRAS461.3494} sample and our $\Delta V_\mathrm{CO}-M_\mathrm{BH}$ correlation (green histogram) and that determined by \protect\cite{Shankar+2009ApJ690.20} from the local AGN luminosity function (solid black line). Error bars are given by the square-root of the sum of the squared weights. The vertical dashed line indicates the SMBH mass corresponding to the $10^{10}\,$M$_\odot$ stellar mass limit of the COLD GASS sample, assuming the correlation of \protect\cite{Beifiori+2012MNRAS419.2497}, below which we expect our sample to be incomplete.}
    \label{fig_SMBH_mass_function}
\end{figure}

The parent COLD GASS sample was selected to be flat in $\log (M_\ast)$, although the need for robust double-horned profiles implies that the sample of \cite{Tiley+2016MNRAS461.3494} does not exactly match this criterion (see the top panel of Figure \ref{fig_SMBH_mass_function}). Nevertheless, we need to weight the sample to match a representative galaxy stellar mass function. We adopt the approach described by \cite{Catinella+2018MNRAS476.875}, whereby we assume the local galaxy stellar mass function of \cite{Baldry+2012MNRAS421.621}, predict the number of galaxies in a volume-limited sample of equal size in $0.2\,$dex stellar mass bins, and weight the SMBH masses predicted from the CO line widths by the ratio of these predictions to the actual number of galaxies in each bin. The estimated SMBH mass function is shown in the lower panel of Figure~\ref{fig_SMBH_mass_function}.

The mass function estimated from our correlation shows good agreement with that estimated by \cite{Shankar+2009ApJ690.20} from the AGN luminosity function assuming a fixed radiative efficiency $\eta=0.065$ and bolometric-to-Eddington luminosity ratio $L_\mathrm{bol}/L_\mathrm{Edd}{=}0.45$. Significant deviations occur at SMBH masses greater than $10^9\,$M$_\odot$, but we note that each populated bin at $M_\mathrm{BH}>10^9\,$M$_\odot$, in addition to the $10^{8.2}-10^{8.4}\,$M$_\odot$ bin that lies substantially below the \cite{Shankar+2009ApJ690.20} results, is based on only a single galaxy and therefore has a large associated uncertainty. In addition, the SMBH masses in the sample used for our $\Delta V_\mathrm{CO}-M_\mathrm{BH}$ correlation poorly sample these most massive SMBHs (see e.g. Figure \ref{fig_VcSMBH}). Furthermore, for $M_\mathrm{BH}>10^{10}\,\mathrm{M_\odot}$, the $M_\mathrm{BH}{-}\sigma_\ast$ relation appears to saturate \citep[e.g.][]{Gultekin+2009ApJ698.198, McConnellMa2013APJ764.184, Krajnovic+2018MNRAS473.5237}. We have no SMBH in this regime in our sample, and so cannot determine whether our correlation continues to hold or not at these masses.

Naturally, estimating a SMBH mass function requires a very careful analysis of potential biases in the underlying sample; that is beyond the scope of this paper. We highlight the morphological biases in the CO detection fraction, and the integration limit of the COLD GASS sample of $M_\mathrm{H_2}/M_\ast=0.015$, as factors we have not controlled for here. Nevertheless, the agreement with the \cite{Shankar+2009ApJ690.20} result is encouraging, particularly since their result arises from accretion physics and a photometric measurement while ours arises from dynamical measurements. 

\section{Conclusions}
\label{sec_conclusion}
CO line emission has previously been used as a tracer of the central parts of galaxy rotation curves, including for the Tully-Fisher relation. The CO discs typically do not extend to halo-dominated radii, and thus in any given galaxy the width of the CO line probes the stellar potential, analogously to the central stellar velocity dispersion $\sigma_\ast$. Although in a LTG one might suppose that the `disc-halo conspiracy' implies that the CO line width measures a flat rotation velocity equivalent to that measured with neutral gas, the same has not been shown for ETGs. The CO line width has however previously been used as a proxy for the stellar velocity dispersion, that is often hard to measure and is strongly affected by both dust extinction and finite apertures.

In this paper, we proposed a correlation between SMBH masses and CO line widths. We investigated this correlation using two samples of galaxies with CO line emission. The first is comprised of galaxies with synthesised spectra from spatially-resolved observations, with generally very high signal-to-noise ratios (SNRs). These spectra were constructed by summing emission within a mask defined from a smoothed version of the original data cube. The second sample is comprised of galaxies with single-dish observations, either from the literature or from new observations conducted at the IRAM 30-m and OSO 20-m telescopes. All the galaxies used have robust dynamical SMBH mass measurements. 

Each CO line width was measured as the FWHM of a profile fit using a Gaussian double peak profile, that has been previously shown to recover well the intrinsic widths of noisy double-horned spectra arising from rotating discs. The line width uncertainties were estimated by Monte Carlo sampling from the determined parameter uncertainties.

We find a good correlation of the CO line widths with both SMBH masses and the central stellar velocity dispersions. There is some evidence that the SMBH mass correlation is steeper for LTGs, without increased scatter. However, the stellar velocity dispersion correlation exhibits a higher intrinsic scatter for LTGs than for ETGs. Using only the spatially-resolved sample yields tighter correlations, and we suggest that the lower SNRs and less robust selection of unresolved observations account for this.

The tightest correlation is found from our spatially-resolved sample as
\begin{equation}
\log\left(\frac{M_\mathrm{BH}}{\mathrm{M_\odot}}\right) = (8.5\pm0.9)\left[\log\left(\frac{W_{50}}{\sin i\,\mathrm{km\,s}^{-1}}\right) -2.7 \right] + (7.5\pm0.1)\,,
\end{equation}
with a total scatter of $0.6\,$dex in the $\log\,M_\mathrm{BH}$ direction, dominated by the intrinsic scatter of $0.5\,$dex.  This intrinsic scatter is comparable to those found for other popular host property--SMBH mass correlations, and it is not dramatically worse than the $0.3\,$dex of intrinsic scatter in the SMBH masses conventionally adopted for the $M_\mathrm{BH}{-}\sigma_\ast$ relation.

We applied our adopted correlation to the CO line widths measured in the COLD GASS survey as part of previous CO Tully-Fisher relation studies, thereby estimating the galaxies' SMBH masses, and constructed a local SMBH mass function, correcting for the original sample's bias in stellar mass. We showed that our SMBH mass function thus derived is consistent with that estimated from the local AGN luminosity function. 

We suggest that our correlation has significant value to estimate SMBH masses where the conventional proxies are unavailable. The CO observations required are simple to make and avoid the need for complicated bulge-disc decompositions. We have further shown that the use of resolved CO observations to generate synthesised spectra dramatically improves the line width measurements. We also suggest that substantial improvements could be made in the CO Tully-Fisher relation by using (low-resolution) interferometric observations, such as those available with the Atacama Compact Array, as these observations allow more robust sample selection and the high SNRs yield significantly smaller line width measurement uncertainties. 

\section*{Acknowledgements}
The authors would like to thank Henrik Olofsson at the Onsala Space Observatory for carrying out observations and Alfred Tiley for providing the COLD GASS Gaussian double peak profile fits used in Section \ref{ssec_application}. We would also like to thank our referee for helpful suggestions.

MDS acknowledges support from a Science and Technology Facilities Council (STFC) DPhil studentship under grant ST/N504233/1. MB was supported by STFC consolidated grant `Astrophysics at Oxford' ST/H002456/1 and ST/K00106X/1. TAD acknowledges support from STFC through grant ST/S00033X/1. MC acknowledges support from a Royal Society University Research Fellowship. 

This paper makes use of the following ALMA data: 
ADS/JAO.ALMA
\#2013.1.00247.S,
\#2015.1.00086.S,
\#2015.1.00419.S,
\#2015.1.00466.S,
\#2015.1.00497.S,
\#2015.1.00598.S,
\#2015.1.00896.S,
\#2015.1.00956.S,
\#2016.1.00437.S,
\#2016.2.00046.S,
\#2016.2.00053.S,
\#2017.1.00301.S,
\#2017.1.00391.S and
\#2018.1.00397.S. 
ALMA is a partnership of ESO (representing its member states), NSF (USA) and NINS (Japan), together with NRC (Canada), MOST and ASIAA (Taiwan) and KASI (Republic of Korea), in cooperation with the Republic of Chile. The Joint ALMA Observatory is operated by ESO, AUI/NRAO and NAOJ.

This publication has received funding from the European Union's Horizon 2020 research and innovation programme under grant agreement No.$\,$730562 [RadioNet]. This research has made use of the NASA/IPAC Extragalactic Database (NED), which is operated by the Jet Propulsion Laboratory, California Institute of Technology, under contract with the National Aeronautics and Space Administration. This paper has also made use of the HyperLeda database (\href{http://leda.univ-lyon1.fr}{http://leda.univ-lyon1.fr}).

\section*{Data Availability}
The data underlying this article are available in Zenodo, at \href{https://dx.doi.org/10.5281/zenodo.4067034}{https://dx.doi.org/10.5281/zenodo.4067034}.




\bibliographystyle{mnras}
\bibliography{./papers} 



\appendix

\section{New Single-dish observations}
\label{app_newObservations}
\subsection{IRAM 30-m telescope observations}
\label{app_191-18}
We carried out new observations of 51 galaxies using the IRAM 30-m telescope on Pico Veleta, Spain, in programme 191-18, between 26th and 31st December 2018. Twenty-two of these galaxies were detected.

The observations used the wobbler to switch between the source and two off-source positions azimuthally-separated from the source by $\pm160\arcsec$, significantly larger than the expected sizes of the sources observed. If there had been emission at these off-source positions, the spectra would exhibit negative features, but there is no such feature.

The heterodyne Eight Mixer Receiver \citep[EMIR;][]{Carter+2012AandA538.89} was used to simultaneously observe the \mbox{$^{12}$CO(1-0)} and $^{12}$CO(2-1) transitions. The receiver was tuned to frequencies allowing several sources at similar redshifts to be observed sequentially without retuning, and the spectra were subsequently corrected to remove the resulting frequency and velocity offsets. Although four sidebands are available at each frequency, only some were recorded due to limitations in the manner backends can be connected.

Strong, compact, continuum sources were observed as pointing and focusing calibrators. At the beginning of each observing shift, at 2-3 hour intervals, after long slews, and after sunrise and sunset, both pointing and focusing calibrations were performed. Additional pointing calibrations were performed before observing each science target, unless consecutive science targets were close by on the sky, using calibrators near the targets. Pointing calibrations were thus performed typically every hour. The calibrations adopted were the means of those required at $115\,$ and $230\,$GHz, except when the calibrations were not well-determined for one of these (typically at $230\,$GHz), when typical offsets from the other frequency (typically $115\,$GHz) were adopted. For these calibrations, we used the Broad Band Continuum (BBC) backend, that covers the entire EMIR band.

Each science source was observed using multiple sets of $12\times30$ second sub-scans, each processed online into a single scan forming the basis of our data reduction. Observations of hot and cold loads for flux calibration were taken every two science scans, and this cycle was repeated until either a strong CO detection was achieved in both lines, or the noise in $T_\mathrm{A}^\ast$ units was less than $3\,$mK. This limit corresponds to $M_\mathrm{H_2}{\approx}10^7\,$M$_\odot$ in a galaxy at the distance of the Virgo cluster, assuming a typical line-of-sight projected line width of $400\,$km$\,$s$^{-1}$.

Two backends were used for science observations: the Fourier Transform Spectrograph \citep[FTS;][]{Klein+2012AandA542.3} and the Wideband Line Multiple Autocorrelator (WILMA\footnote{\href{http://www.iram.fr/IRAMFR/TA/backend/veleta/wilma/index.htm}{http://www.iram.fr/IRAMFR/TA/backend/veleta/wilma/index.htm}}). FTS was used in the `wide' mode, with eight units each with a bandwidth of ${\approx}4\,$GHz and a channel width of ${\approx}200\,$kHz, corresponding to total bandwidths of ${\approx}\mbox{10\,400}\,$ and ${\approx}5200\,$km$\,$s$^{-1}$, and channel widths of ${\approx}0.5\,$ and ${\approx}0.25\,$km$\,$s$^{-1}$, at $115\,$ and $230\,$GHz, respectively. Four FTS units were attached to the $^{12}$CO(1-0) sideband and two to the $^{12}$CO(2-1) sideband. The remaining two FTS units were attached to the upper outer sideband on the $230\,$GHz receiver, as required by the connector box, although no emission line was expected within this sideband. Each pair of FTS units measured the horizontal and vertical polarisations of the band separately, and these were averaged together.
WILMA served as a backup with a $16\,$GHz bandwidth and a $2\,$MHz channel width. 

Data reduction was performed off-line using the \texttt{Continuum and Line Analysis Single-dish Software (CLASS)} from the \texttt{Grenoble Image and Line Data Analysis Software (GILDAS)} suite\footnote{\href{http://www.iram.fr/IRAMFR/GILDAS}{http://www.iram.fr/IRAMFR/GILDAS}}. Every 6-minute scan was manually checked for bad data. The frequency axis was transformed from that used to tune the receiver for groups of multiple targets to the true rest-frequencies of the $^{12}$CO transitions and then to velocities using the optical convention. FTS operates on three sections of each spectrum, which can lead to `platforming' errors, i.e. absolute flux offsets between the sections. This effect was removed for every 6-minute scan by masking detected lines and subtracting a first-order fit to the remaining baseline in each section. We then averaged together all polarisations and (good) scans for each science source and binned to $30\,$km$\,$s$^{-1}$ channels. We performed another fit to the baseline to remove any remaining offset arising from the background, again masking regions where the CO line was detected. Additionally, as many of these galaxies are sufficiently local that the $^{12}$CO(1-0) line lies close to the $118\,$GHz atmospheric oxygen line, the high-frequency channels were also masked where they exhibited increased noise. Although the bandpass never contained the emission line itself, it often included its broad wings.

Many of our sources were observed at relatively low elevations, of $20$-$30\degree$. In addition to the increased airmass at low elevations, and hence the longer integrations required to reach the same sensitivity, the 30-m telescope exhibits an elevation-dependent gain due to small surface deformations \citep[see e.g.][]{Greve+1998AandAS132.413}. For a point source, the peak gain is found at $50\degree$ elevation, whereas at $20\degree$ it has dropped to $92\%$ at $145\,$GHz and $80\%$ at $210\,$GHz. For an extended source, and at lower frequencies, this correction is expected to be less significant. We therefore assume that we can neglect this correction for all sources. Nevertheless, the fluxes adopted may be underestimated by up to $\approx20\%$ for a compact source at $^{12}$CO(2-1).

We do not know \textit{a priori} the extent of the CO emission in any source. However, since the telescope does not have a uniform sensitivity across the beam, and our targets are expected to be extended, we should in principle correct for the coupling between the source emission distribution and the telescope's beam,  as expressed through the $K$-factor \citep{Baars1973ITAP21.461}. For a circular Gaussian source as extended as the beam, the correction can be as large as a factor of 2, whereas for a point-source, no correction is required. For simplicity, and having noted the danger of assuming any particular source shape, we neglect this $K$-factor here, implicitly assuming that all sources are point-like. This decision is intended to maximise the utility of our data to the community, rather than imposing our choice of assumptions. As noted in the main text, none of these corrections would affect our measured line widths.

\begin{figure*}
	\captionsetup[subfloat]{captionskip=10pt}
		\subfloat[ARK 120]{\includegraphics[trim={0.5cm 1cm 1.5cm 1cm}, width=0.3\textwidth]{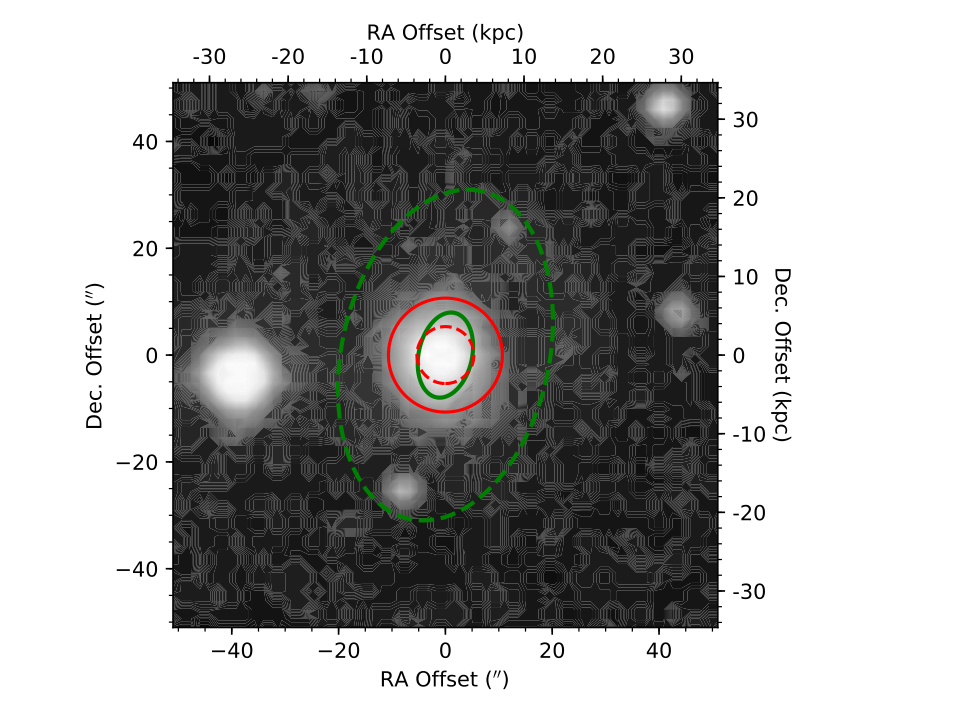}
		\includegraphics[trim={0.5cm 1cm 1.5cm 1cm}, width=0.3\textwidth]{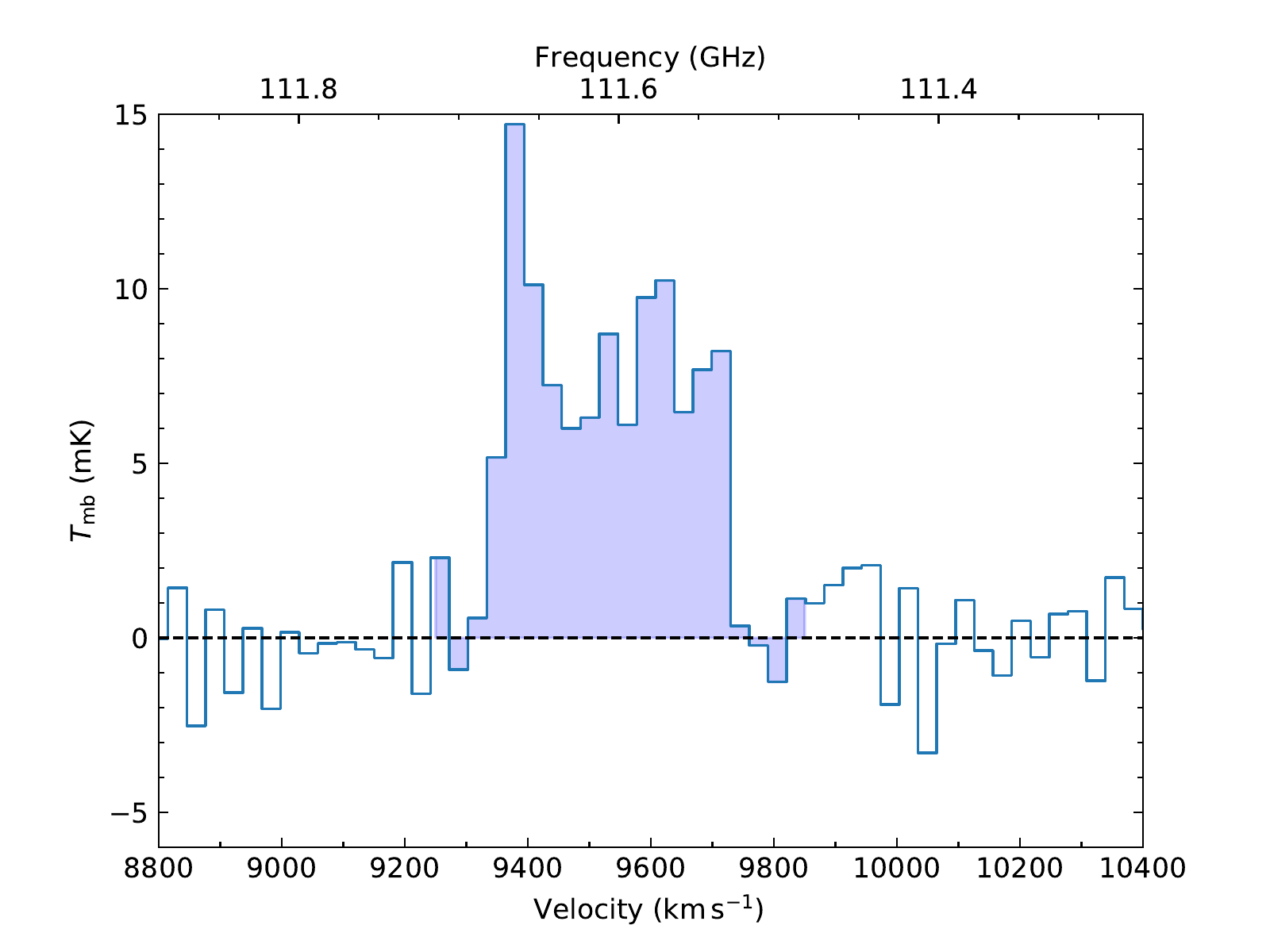} 
		\includegraphics[trim={0.5cm 1cm 1.5cm 1cm}, width=0.3\textwidth]{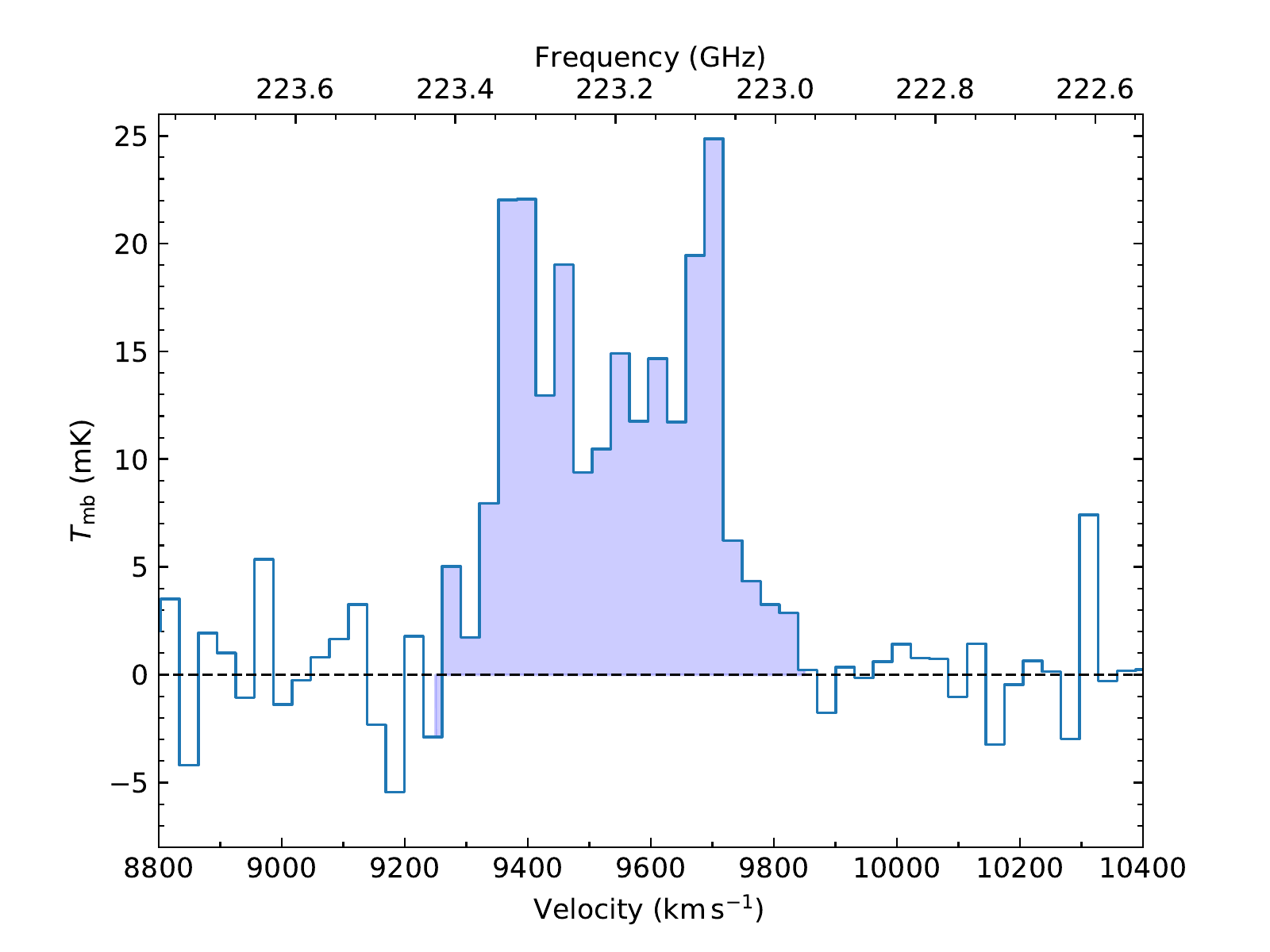} 
		}
		
    \caption{One of the 22 galaxies detected in our CO observations using the IRAM 30-m telescope (programme 191-18). For each galaxy, the left panel shows the extent of the 30-m beam at $^{12}$CO(1-0) (red solid circle) and $^{12}$CO(2-1) (red dashed circle), overlaid on an image of the galaxy from DSS. Also overlaid are the areas enclosed by one effective radius (from \protect\citealt{vdBosch2016ApJ831.134}; green solid ellipse) and $D_\mathrm{25}$ (from \href{http://leda.univ-lyon1.fr}{HyperLEDA}; green dashed ellipse), adopting the inclination and position angle listed in \href{http://leda.univ-lyon1.fr}{HyperLEDA} and the distance from \protect\cite{vdBosch2016ApJ831.134}. The central and right panels show the spectrum of the $^{12}$CO(1-0) and $^{12}$CO(2-1) emission line, respectively. The velocity range used to determine the integrated flux is shaded. All 22 detected galaxies are shown in Figure A3 in the supplemental material.}
    \label{fig_191-18spectra}
\end{figure*}

Table \ref{tab_191-18Results} describes the spectra obtained for each galaxy. The detected lines are shown in Figure \ref{fig_191-18spectra}. For the same reasons as above, the associated quantities listed are in main beam temperature units, without $K$-factor correction. The fluxes listed are thus formally lower limits, but they are likely only slightly underestimated. We convert to main beam temperatures using the 30-m telescope efficiencies listed in Table \ref{tab_telEff}.

Integrated line fluxes are measured as the integral over the velocity range listed in Table \ref{tab_191-18Results} and indicated by shading in Figure \ref{fig_191-18spectra}, i.e.
\begin{equation}
\label{eq_intFlux}
I = \int^{v_1}_{v_0} T_\mathrm{mb}(v) dv\,,
\end{equation}
where $T_\mathrm{mb}(v)$ is the flux in each velocity channel (of width $dv$) expressed as a main beam temperature, and the integral is taken over the velocity range $v_0$ to $v_1$.

We adopt the standard estimate of the uncertainty in the integrated flux \citep[e.g.][]{Sage+2007ApJ657.232,Young+2011MNRAS414.940}, 
\begin{equation}
\sigma = \sigma_\mathrm{rms}\,\Delta v \sqrt{N_\mathrm{line}\left(1+\frac{N_\mathrm{line}}{N_\mathrm{noise}}\right)}\,,
\label{eq_intNoise}
\end{equation}
where $ \sigma_\mathrm{rms}$ is the noise per channel (measured in line-free channels), $\Delta v$ the velocity width of each channel, $N_\mathrm{line}$ the number of channels integrated over, and $N_\mathrm{noise}$ the number of channels used to estimate the noise, that we assume is approximately the number of channels in the bandpass.
The line ratios are thus $I_{(2-1)}/I_{(1-0)}$ in main beam temperature units. Molecular gas masses for detected lines are estimated adopting \mbox{$X_\mathrm{CO}\,=\,2\,\times\,10^{20}\,$mol$\,$cm$^{-2}\,$(K$\,$km$\,$s$^{-1}$)$^{-1}$}, and the quoted uncertainties include the uncertainties in the distances adopted from \cite{vdBosch2016ApJ831.134}. For each undetected galaxy, we estimate an upper limit on the molecular gas mass by assuming the line width predicted from the CO Tully-Fisher Relation \citep{Tiley+2016MNRAS461.3494}, calculating the sensitivity using Equation \ref{eq_intNoise}, and hence the H$_2$ mass that would be (marginally) detected at $1\sigma$ sensitivity using Equation \ref{eq_intFlux}.

\begin{table*}
\begin{center}
\caption{New observations using the IRAM 30-m telescope.}
\label{tab_191-18Results}
\begin{tabular}{lccccccccl}
    \hline
    Galaxy & T-type & rms (1-0) &  rms (2-1) & Velocity range & $I_{(1-0)}$ & $I_{(2-1)}$ & $\frac{I_{(2-1)}}{I_{(1-0)}}$ & $\log\left(\frac{M_{\mathrm{H_2}}}{\mathrm{M_\odot}}\right)$ & Notes\\ 
                & & (mK) &  (mK) & (km$\,$s$^{-1}$) & (K km$\,$s$^{-1}$) & (K km$\,$s$^{-1}$) &  & & \\ 
         (1) & (2) & (3) & (4) & (5) & (6) & (7) & (8) & (9) & (10)\\
    \hline
    3C382 & \phantom{1}-5.0 & 1.8 & 2.9 & & & & &$<8.9$ \\
    ABELL 1836 BCG & \phantom{1}-3.2 &1.7 & 1.8 & & & & & $<8.3$ &\\
    ARK 120 & \phantom{1}-5.0 & 1.5 & 2.4 & $9250-9850$ & \phantom{0}$3.2\pm0.2$ &  \phantom{00}$6.9\pm0.3$ & 2.1 & $9.4\pm0.1$ \\
    ESO 558-009 & \phantom{-1}3.9& 3.1 & 4.4 & $7000-7900$ &  \phantom{0}$5.0\pm0.5$ &  \phantom{00}$9.2\pm0.8$ & 1.8 & $9.3\pm0.1$ &\\
    H0507+164 &--& 2.4 & 3.7 & $5000-5500$ &  \phantom{0}$1.7\pm0.3$ &  \phantom{00}$3.6\pm0.5$ & 2.1 & $8.6\pm0.1$& \\
    IC 1481 & \phantom{-1}8.0 & 2.2 & 9.5 & $5800-6400$ &  \phantom{0}$5.2\pm0.3$ &  \phantom{0}$10.5\pm1.4$ & 2.0 & $9.2\pm0.1$ & (1)\\
    J0437+2456 & \phantom{-}10.0 &1.8 & 2.7 & $4600-5100$ &  \phantom{0}$3.1\pm0.2$ &  \phantom{00}$3.1\pm0.4$ & 1.0 & $8.7\pm0.1$ &\\
    Mrk 279 &\phantom{1}-2.0 & 1.5 & 2.7 & & & & & $<8.2$\\
    Mrk 1029 & -- &2.1 & 2.7 & $8700-9200$ &  \phantom{0}$4.9\pm0.3$ &  \phantom{0}$14.8\pm0.3$ & 3.0& $9.5\pm0.1$ \\
    NGC 193 & \phantom{1}-3.0 & 2.1 & 3.3 & & & & & $<7.3$  \\
    NGC 307 & \phantom{1}-1.9 & 2.1 & 3.3 & & & & & $<7.4$\\
    NGC 547 & \phantom{1}-4.8 & 2.7 & 2.9 & & & & & $<7.7$& \\
    NGC 613 & \phantom{-1}4.0 & 6.8 & 4.3 & $1200-1800$ & $87.3\pm0.9$ & $137.8\pm0.6$ & 1.6 & $8.9\pm0.1$ & (1)\\
    NGC 1271 & \phantom{1}-2.0 & 1.8 & 2.9  & & & & & $<7.8$ &  \\
    NGC 1300 & \phantom{-1}4.0 & 5.9 & 4.6 & $1300-1800$ & $28.6\pm0.8$ &  \phantom{0}$45.1\pm0.6$ & 1.6  & $8.7\pm0.5$ & (1) \\
    NGC 1398 & \phantom{-1}2.0 & 3.2 & 3.5 & & & & & $<6.9$ &  \\
    NGC 1497 & \phantom{1}-2.0 & 1.2 & 2.2 & $5800-6500$ &  \phantom{0}$2.3\pm0.1$ &  \phantom{00}$2.0\pm0.3$ & 0.8 & $8.7\pm0.1$ & \\
    NGC 1550 & \phantom{1}-4.1 & 2.3 & 2.2 & & & & & $<7.4$ \\
    NGC 1600 & \phantom{1}-4.6 & 2.6 & 3.0 & & & & & $<7.7$ & \\
    NGC 1961 & \phantom{-1}4.2 & 6.0 & 4.1 & $3500-4300$ & $90.6\pm1.0$& $134.2\pm0.6$ & 1.5 & $9.9\pm0.1$ & (1) \\
    NGC 2179 & \phantom{-1}0.1 & 2.9 & 2.5 & $2650-3200$ & $60.3\pm0.4$ &  $153.4\pm0.3$ & 2.5 & $8.0\pm0.1$ &  \\
    NGC 2748 & \phantom{-1}4.0 & 10.5\phantom{0} & 9.4 & $1400-1700$ & $20.0\pm1.0$ &  \phantom{0}$26.9\pm1.0$ & 1.3 & $8.6\pm0.4$& (1) \\
    NGC 2787 & \phantom{1}-1.0 & 5.0 & 4.1 & & & & & $<5.9$&  \\
    NGC 2911 & \phantom{1}-2.0 & 1.9 & 2.4 & $2900-3600$ &  \phantom{0}$2.9\pm0.3$ &  \phantom{00}$4.9\pm0.3$ & 1.7 & $8.3\pm0.1$ & \\
    NGC 2960 & \phantom{-1}0.8 & 2.9 & 3.5 & $4650-5200$ &  \phantom{0}$7.0\pm0.4$ &  \phantom{0}$12.9\pm0.5$ & 1.8 & $9.1\pm0.1$ & \\
    NGC 4151 & \phantom{-1}1.9 & 1.9 & 2.2 & \phantom{0}$900-1200$ &  \phantom{0}$2.8\pm0.1$ &  \phantom{00}$3.3\pm0.2$ & 1.2 & $7.6\pm0.1$ & \\
    NGC 4291 & \phantom{1}-4.8 & 2.6 & 3.0 & & & & & $<6.8$ & \\
    NGC 4335 & \phantom{1}-4.4 & 2.2 & 3.3 & & & & & $<7.5$ & (2) $v_0=4050\,$km$\,$s$^{-1}$\\
                      & & 2.7 & 4.6 & & & & & $<7.6$ & (2) $v_0=500\,$km$\,$s$^{-1}$\\
    NGC 4350 & \phantom{1}-1.8 & 2.4 & 3.0 & & & & & $<6.4$ &  \\
    NGC 4371 & \phantom{1}-1.3 & 2.3 & 2.5 & & & & & $<6.3$ & \\
    NGC 4486A & \phantom{1}-5.0 & 2.7 & 2.7 & & & & & $<6.3$ &  \\
    NGC 4486B & \phantom{1}-5.0 & 2.2 & 2.7 & & & & & $<6.2$ &  \\
    NGC 4698 & \phantom{-1}1.6 & 2.8 & 3.2 & & & & & $<6.4$ & \\
    NGC 4889 & \phantom{1}-4.3 & 1.8 & 2.4 & & & & & $<8.0$ & \\
    NGC 5077 & \phantom{1}-4.8 & 2.9 & 3.2 & & & & & $<7.3$ & \\
    NGC 5252 & \phantom{1}-2.0 & 2.1 & 2.9 & & & & & $<8.2$ & \\
    NGC 5328 & \phantom{1}-4.7 & 2.8 & 3.2 & & & & & $<7.8$ & \\
    NGC 6086 & \phantom{-1}4.0 & 2.2 & 3.7 & & & & & $<8.4$ & \\
    NGC 6251 & \phantom{1}-4.9 & 3.1 & 4.1 & & & & & $<8.3$ & \\
    NGC 6264 & \phantom{-1}2.7 & 1.7 & 2.5 & $\phantom{0}9700-10200$ &  \phantom{0}$1.4\pm0.3$ &  \phantom{00}$1.0\pm0.3$ & 0.7 & $9.1\pm0.1$\\
    NGC 6323 & \phantom{-1}2.0 & 1.8 & 2.4 & $7400-8000$ &  \phantom{0}$4.2\pm0.3$ &  \phantom{00}$4.3\pm0.3$ & 1.0  & $9.3\pm0.1$ \\
    NGC 6500 & \phantom{-1}1.7 & 2.7 & 2.9 & $2800-3400$ &  \phantom{0}$9.2\pm0.4$ &  \phantom{00}$9.8\pm0.5$ & 1.1 & $8.7\pm0.1$ \\
    NGC 7619 & \phantom{1}-4.8 & 2.1 & 2.9 & & & & & $<7.4$ & \\
    NGC 7626 & \phantom{1}-4.8 & 2.3 & 2.9 & & & & & $<7.1$ & \\
    NGC 7682 & \phantom{-1}1.6 & 1.8 & 4.8 & $4900-5200$ &  \phantom{0}$1.9\pm0.1$ &  \phantom{00}$3.9\pm0.5$ & 2.6 & $8.4\pm0.1$\\
    NGC 7768 & \phantom{1}-4.8 & 2.3 & 3.3 & & & & & $<8.2$ & \\
    UGC 1214 & \phantom{1}-1.2 & 2.4 & 2.7 & $4900-5300$ &  \phantom{0}$2.1\pm0.3$ &  \phantom{00}$2.9\pm0.3$ & 1.4 & $8.5\pm0.1$ \\
    UGC 1841 & \phantom{1}-4.9 & 2.2 & 3.5 & & & & & $<7.7$ & \\
    UGC 3789 & \phantom{-1}1.6 & 4.2 & 3.7 & $3000-3500$ & $11.7\pm0.5$ &  \phantom{0}$13.9\pm0.5$ & 1.2 & $9.1\pm0.1$ & (1) \\
    UGC 6093 & \phantom{-1}3.7 & 1.9 & 3.2 & $10\,400-10\,600$ &  \phantom{0}$2.1\pm0.1$ &  \phantom{00}$1.5\pm0.3$ & 0.7 & $9.3\pm0.1$ & \\
    UGC 12064 &\phantom{1}-3.0 & 2.2 & 3.5 & & & & & $<7.0$ & \\
    \hline
    \end{tabular}
\end{center}
\parbox[t]{\textwidth}{\textbf{Notes:} Column 1 lists the name of each target galaxy, as listed in Table 2 of \protect\cite{vdBosch2016ApJ831.134}. The morphological classification on the numerical Hubble scale from \href{http://leda.univ-lyon1.fr}{HyperLEDA} is listed in Column 2. Columns 3 and 4 indicate the rms noise achieved in the sideband containing the CO(1-0) and CO(2-1) line, respectively. For detected galaxies, columns 5-8 list the velocity range used in Equation \ref{eq_intFlux}, the integrated CO(1-0) and CO(2-1) flux, and the associated line ratio, respectively. All fluxes are expressed as main beam temperatures ($T_\mathrm{mb}$). Column 9 contains the estimated total molecular gas mass or upper-limit, calculated as described in the text. Finally, Column 10 contains notes on specific galaxies as follows. (1) This source was strongly detected in both lines before the intended sensitivity was reached, so the noise exceeds $3\,$mK in at least one line. (2) NGC4335 was initially not detected at the $4631\,$km$\,$s$^{-1}$ systemic velocity listed in the \href{http://ned.ipac.caltech.edu}{NASA/IPAC Extragalactic Database}. This was surprising due to the prominent dust disc visible in \textit{HST} images. The Sloan Digital Sky Survey lists a velocity of $-111\,$km$\,$s$^{-1}$, so we obtained a second spectrum centred on the corresponding frequency. The galaxy remained undetected. The velocities given in the table are the centres of the band for each of these two spectra.}
\end{table*}

\subsection{Onsala Space Observatory 20-m observations}
\label{app_2018-04a}
A few of the most extended targets were observed using the Onsala Space Observatory (OSO) 20-m telescope in service mode, largely through an automated script. The observations were carried out in two observing sessions 20th-23rd November 2018 and 27th Februrary - 6th March 2019, using the $3\,$mm receiver \citep{Belitsky+2015AandA580.A29} to cover the $^{12}$CO(1-0) line. This yields a $4\,$GHz bandpass, that is fully-sampled by the Omnisys A (OSA) backends with a $76\,$kHz raw channel width. At $115\,$GHz, this is equivalent to a bandpass of ${\approx}10\,000\,$km$\,$s$^{-1}$ and a channel width of ${\approx}0.2\,$km$\,$s$^{-1}$. The two backend units cover the horizontal and vertical polarisations simultaneously, that are then co-added to improve the SNR. The observations were carried out in dual beam-switching mode, whereby the source is moved between the signal beam and the reference beam, and the two spectra are subtracted from each other. This beam-switching mode has a fixed $11\arcmin$ throw.

For each source, individual scans of 4 minutes each were manually inspected, and those with bad baselines removed. This issue significantly affected the observations of NGC~4335, for which around half the scans had to be eliminated. Using \texttt{CLASS}, the remaining scans were then averaged for each source and binned to $30\,$km$\,$s$^{-1}$ channels. The detected emission lines and the region of increased noise close to the $118\,$ GHz atmospheric oxygen line were then masked, and a linear baseline fit carried out and subtracted. Four of the 9 sources observed were detected in $^{12}$CO(1-0).

The native $T_\mathrm{A}^\ast$ units were converted to $T_\mathrm{mb}$ using the main beam efficiencies calculated for each scan. The spectra were then integrated over the line to calculate the integrated flux and corresponding molecular gas mass, following the prescription given in Section \ref{app_191-18} (Equations \ref{eq_intFlux} and \ref{eq_intNoise}). The results are listed in Table \ref{tab_o2018-04aResults} and the spectra of the detected sources are shown in Figure \ref{fig_o2018-04a_spectra}. 

The conversion factor from temperature to flux units is not available for $115\,$GHz, but we rescaled the value given at $86\,$GHz, assuming that the product of the conversion factor and main beam efficiency is constant, as suggested by OSO staff. Thus, 
\begin{equation}
S\mathrm{(Jy)} = \frac{10.7\,\mathrm{Jy\,K^{-1}}}{\eta_\mathrm{mb}} T_\mathrm{A}^\ast\,,
\end{equation}
where we have assumed a point source.

\begin{table*}
\begin{center}
\caption{New observations using the Onsala Space Observatory 20-m telescope.}
\label{tab_o2018-04aResults}
\begin{tabular}{lcccccl}
    \hline
    Galaxy & T-type & rms (1-0)  & Velocity range & $I_{(1-0)}$  & $\log\left(\frac{M_{\mathrm{H_2}}}{\mathrm{M_\odot}}\right)$ & Notes\\ 
                & & (mK) & (km$\,$s$^{-1}$) & (K km$\,$s$^{-1}$) & & \\ 
           (1) & (2) & (3) & (4) & (5) & (6) & (7)\\
    \hline
    Mrk 279 & -2.0 & $\phantom{1}1.8$ & $8850-9200$ & $1.0\pm0.2$ & $8.8\pm0.1$ \\
    NGC 2273 & \phantom{-}0.9 & $\phantom{1}5.0$ & $1600-2200$ & $8.8\pm0.7$ & $8.5\pm0.1$  \\
    NGC 2787 & -1.0 & $\phantom{1}5.1$ & & & $<5.9$\\
    NGC 2911 & -2.0 & $\phantom{1}4.3$ & & & $<7.5$\\
    NGC 3079 & \phantom{-}6.4 & $18.0$ & $\phantom{0}800-1600$ & $115\pm3\phantom{.0}$ & $9.1\pm0.1$ & (1)\\
    NGC 4335 & -4.4 & $11.9$ & & & $<8.3$ & (2)\\
    NGC 6251 & -4.9 & $\phantom{1}1.6$ & & & $<8.0$\\
    NGC 7768 & -4.8 & $\phantom{1}1.6$ & & & $<8.0$\\
    UGC 3789 & \phantom{-}1.6& $\phantom{1}4.1$ & $3000-3400$ & $6.1\pm0.5$ & $8.8\pm0.1$\\
    \hline
    \end{tabular}
\end{center}
\parbox[t]{\textwidth}{\textbf{Notes:}  Column 1 lists the name of each target galaxy, as listed in Table 2 of \protect\cite{vdBosch2016ApJ831.134}. The morphological classification on the numerical Hubble scale from \href{http://leda.univ-lyon1.fr}{HyperLEDA} is listed in Column 2. Column 3 indicates the rms noise achieved in the bandpass containing the CO(1-0) line. For detected galaxies, Columns 4 and 5 list the velocity range used in Equation \ref{eq_intFlux}, and the integrated CO(1-0) flux respectively. All fluxes are expressed as main beam temperatures ($T_\mathrm{mb}$). Column 6 contains the estimated total molecular gas mass or upper-limit, calculated as described in the text. Finally, Column 7 contains notes on specific galaxies as follows. (1) $^{12}$CO(1-0) emission was strongly detected in NGC~3079 before the anticipated sensitivity was reached, so the rms noise appears comparatively high. (2) Bad baselines in many of the scans of NGC~4335 yielded a noise higher than requested.}
\end{table*}

\setcounter{figure}{1}
\begin{figure*}
	\captionsetup[subfloat]{captionskip=10pt}
		\subfloat[Mrk 279]{\includegraphics[trim={0.5cm 1cm 1.5cm 1cm}, width=0.3\textwidth]{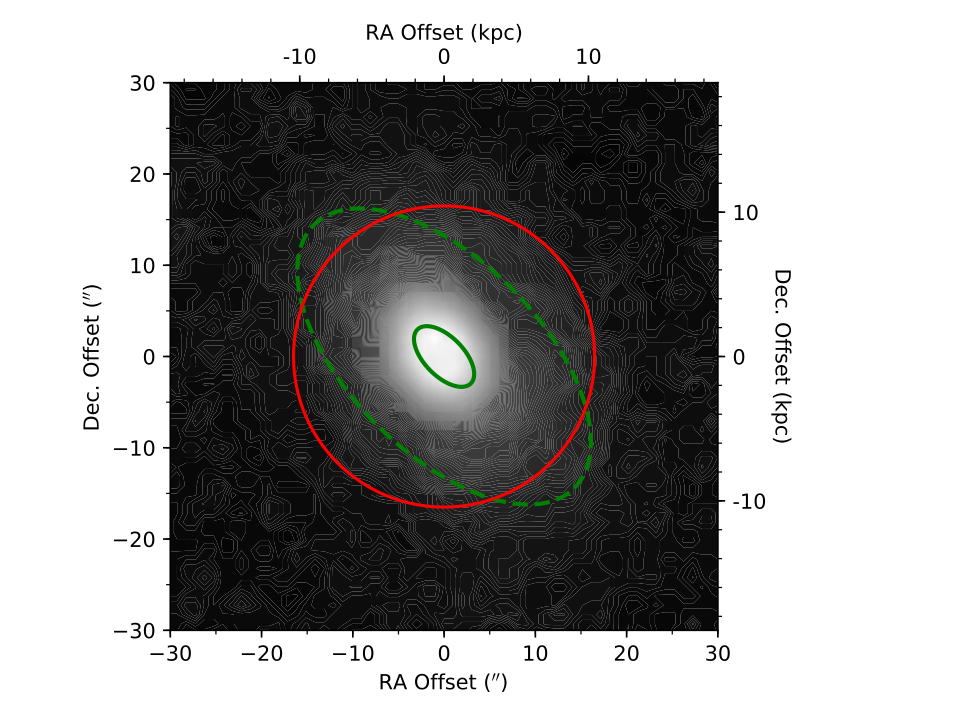}
		\includegraphics[trim={0.5cm 1cm 1.5cm 1cm}, width=0.3\textwidth]{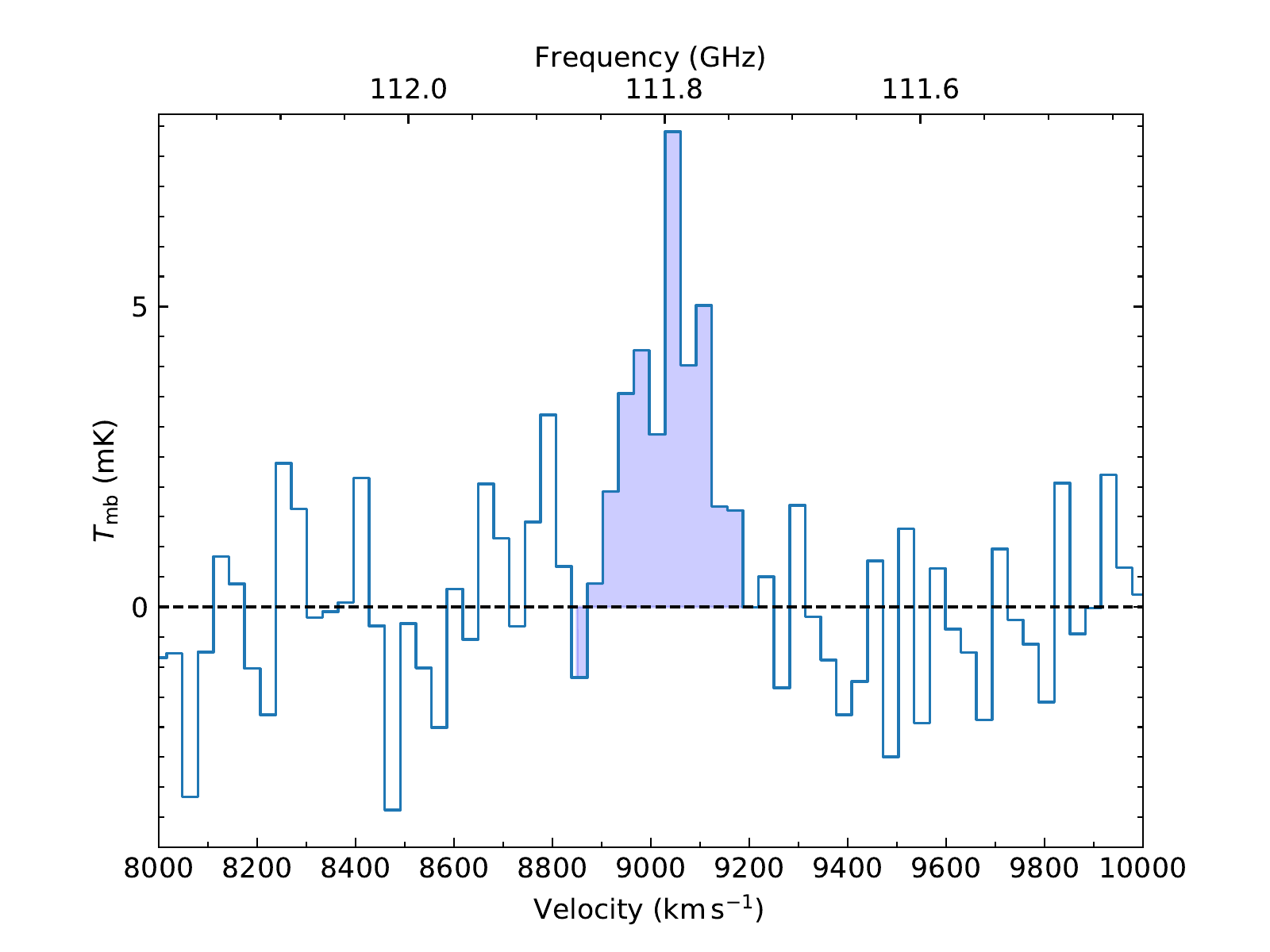} 
		}
		
    \caption{One of the four galaxies detected in our CO observations using the OSO 20-m telescope (programme 2018-04a). For each galaxy, the left panel shows the extent of the 20-m beam at $^{12}$CO(1-0) (red solid circle), overlaid on an image of the galaxy from DSS. Also overlaid are the areas enclosed by one effective radius (from \protect\citealt{vdBosch2016ApJ831.134}; green solid ellipse) and $D_\mathrm{25}$ (from \href{http://leda.univ-lyon1.fr}{HyperLEDA}; green dashed ellipse), adopting the inclination and position angle listed in \href{http://leda.univ-lyon1.fr}{HyperLEDA} and the distance from \protect\cite{vdBosch2016ApJ831.134}. The right panel shows the spectrum of the $^{12}$CO(1-0) emission line. The velocity range used to determine the integrated flux is shaded. All four detections are shown in Figure A4 in the supplementary material.}
   \label{fig_o2018-04a_spectra}
\end{figure*}


\bsp	
\label{lastpage}
\end{document}